# Light Nuclides Produced in the Proton-Induced Spallation of $^{238}$U at 1 GeV

M. V. Ricciardi[(1)], P. Armbruster[(1)], J. Benlliure[(2)], M. Bernas[(3)], A. Boudard[(4)],
S. Czajkowski[(5)], T. Enqvist[(1)], A. Kelić[(1)], S. Leray[(4)], R. Legrain[(4)], B. Mustapha[(3)], J. Pereira[(2)],
F. Rejmund[(1)], K. -H. Schmidt[(1)], C. Stéphan[(3)], L. Tassan-Got[(3)], C. Volant[(4)], O. Yordanov[(1)]

(1) GSI, Planckstr. 1, 64291 Darmstadt - Germany
(2) Univ. Santiago de Compostela, 15706 Santiago de Compostela - Spain
(3) IPN Orsay / IN2P3, 91406 Orsay Cedex - France
(4) DAPNIA / SPhN CEA / Saclay, 91191 Gif sur Yvette Cedex - France
(5) CEN Bordeaux-Gradignan, 33175, Gradignan Cedex- France

**Abstract**

The production of light and intermediate-mass nuclides formed in the reaction $^1$H+$^{238}$U at 1 GeV was measured at the Fragment Separator (FRS) at GSI, Darmstadt. The experiment was performed in inverse kinematics, shooting a 1 $A$ GeV $^{238}$U beam on a thin liquid-hydrogen target. 254 isotopes of all elements in the range $7 \leq Z \leq 37$ were unambiguously identified, and the velocity distributions of the produced nuclides were determined with high precision. The results show that the nuclides are produced in a very asymmetric binary decay of heavy nuclei originating from the spallation of uranium. All the features of the produced nuclides merge with the characteristics of the fission products as their mass increases.

## 1. Introduction

In 1996, at GSI, Darmstadt, a European collaboration started a dedicated experimental program, devoted to reaching a full comprehension of the proton-induced spallation reactions. The accurate knowledge of proton-induced spallation reactions is relevant both for fundamental research and for technical applications. Among the latter, the design of accelerator-driven systems (ADS) and radioactive ion-beam facilities (RIB) relies strongly on the knowledge of the formation cross sections of residual nuclei produced in such reactions. This information is needed to calculate the short-term and long-term radioactivity, building up in these facilities, and thus for designing the shielding and estimating the residual activation of such devices. In ISOL-type radioactive ion-beam facilities, the formation cross sections are decisive to determine which nuclides far from stability can become accessible, and to estimate the attainable secondary-beam intensities, once appropriate extraction and ionisation procedures will be developed. Here, fission is of special interest, because it seems to be best suited for approaching the neutron drip line in the medium-mass range. The energy of 1 GeV/nucleon is estimated to be optimum for both applications [1, 2].

In the past, the available experimental data on spallation reactions were scarce and fragmentary, and the predictive power of the computational codes traditionally used for the design and for the shielding of nuclear facilities was in most cases rather poor [3]. For this reason, a project devoted to studying, understanding and modelling these nuclear reactions at energies around 1 GeV per nucleon started at GSI nine years ago. Within this project, the first comprehensive survey on nuclide production cross sections in $^{197}$Au + $^{1}$H [4, 5] and $^{208}$Pb + $^{1,2}$H [6, 7, 8] at around 1 GeV per nucleon was obtained. Experiments on other systems or at other energies (in the range 0.3-1.5 GeV per nucleon) have been published or are still being analysed ($^{208}$Pb + $^{1}$H [9, 10], $^{238}$U + $^{1,2}$H [11, 12, 13, 14, 15], $^{56}$Fe + $^{1}$H [16,17], and $^{136}$Xe + $^{1}$H [17]). The essential goal of this project is the measurement of the formation cross sections of residual nuclei in few *key* nuclear reactions. From nuclear-reaction theory and from phenomenological observations it is expected that the cross sections for proton-induced reactions above some tens of MeV behave smoothly with target mass and projectile energy. Therefore, three nuclides − $^{56}$Fe, $^{208}$Pb and $^{238}$U − which represent a typical construction material, a target material of the spallation neutron source and a highly fissile nucleus, respectively, were chosen as *key* nuclei to be investigated.

The study of light-residue production (from $Z$=7 to $Z$=37) in hydrogen-induced reactions of $^{238}$U at 1 GeV, presented here, belongs to this systematic study. Together with other measurements, performed in the same experiment, which were dedicated to the formation of heavier residues by fission (from $Z$=28 to $Z$=73 [12,13]) and by spallation-evaporation (from $Z$=74 to $Z$=92 [11]) in the system $^{238}$U + hydrogen, the whole chart of the nuclides from $Z$=7 on was covered. The subdivision of the experiment in different measurements is due to the fact that different experimental techniques and analysis methods were applied in these different mass regions. For example, the heaviest residues could only be identified by the use of a thick energy degrader in the intermediate image plane of the fragment separator [18]. Special conditions were also met in the present work, since a very large range in magnetic rigidity had to be covered for investigating the light fragments. Along with the formation cross-sections, the longitudinal momenta of the fragments after the reaction were also measured. They provide information on the reaction mechanism and are used for revealing the binary (or fission-like) character of the decay due to the Coulomb repulsion between the two fragments. The knowledge on the reaction mechanism is of great importance for the design of RIB facilities and ADS, and in general for the new generation of accelerators at high intensities [19], because the radiation damage of the structures depends on the kinetic energy of the fragments. An overview of all the results for the reaction $^{238}$U on hydrogen at 1 $A$ GeV are given in a dedicated letter [20].

In the context of the production of exotic nuclei, it should be mentioned that 253 very neutron-rich light reaction products in the range $9 \leq Z \leq 46$ from the interaction of $^{238}$U projectiles with beryllium and lead targets were observed in a previous experiment [21] in the course of an experimental programme dedicated to produce new neutron-rich isotopes [22], in particular $^{78}$Ni [23], using the same installations as the present experiment. In contrast, the present experiment emphasises the reaction aspect and focuses on a systematic overview of the most strongly produced isotopes of the light elements in the system $^{238}$U + $^{1}$H.

Apart from the technical applications, the measurement of formation cross sections has also interest for fundamental research. In astrophysics, for instance, they enter into the description of the processes that affect the composition of energetic nuclei during their transport through the Galaxy, from their source to the Earth where they are observed. The models for the propagation of cosmic rays rely heavily on the knowledge of the formation cross sections of light nuclei from the interactions of the heavy nuclei in the interstellar medium [24], which mostly consists of hydrogen. Also the study of the reaction mechanisms responsible for the production of the light nuclides is of great physical interest. In the reaction $^{238}$U + hydrogen at 1 $A$ GeV, apart from spallation reactions, which end up in rather heavy fragments (at $Z \geq 75$, see ref. [11]), most part of the cross section of the medium-mass residues results from fission reactions ([12, 25, 26, 27, 28]). One of the most important signatures of fission is the binary nature of the decay process. The light residues, investigated in this work, also showed a binary nature, but binary decay can occur also in multifragmentation-type processes [29]. The possible scenarios behind fission and multifragmentation are indeed strongly different, because the first presupposes the slow decay of a compound nucleus, while the second one the passage through a fast break-up phase. It was discussed, if the yield of such binary products [30] and the longitudinal momentum transferred to the decaying nucleus [31] can carry information on the reaction mechanism that produced them. In this work, we will discuss whether the light residues that we observed are consistent with one or the other picture, making use of the two available observables, the velocities and the production cross sections of the residues.

**2. The experimental technique**

The experiment was performed in inverse kinematics at relativistic energy, i.e. shooting a 1 $A$ GeV $^{238}$U beam into a H$_2$ target. In these experimental conditions, the fragment escapes the target strongly focussed in forward direction and is detected in-flight prior to its β decay. Thus, the whole isotopic distribution can be obtained for every element, and the velocity of the identified nucleus can be precisely determined and used to deduce the reaction mechanism which generated that isotope. In this way, fission and fragmentation events can be disentangled, as we will show. Another attractive peculiarity of this technique at around 1 $A$ GeV is that the products are fully stripped and can be identified without ambiguity caused by different charge states.

The SIS18 heavy-ion accelerator of GSI, Darmstadt, was used to provide the $^{238}$U beam of 1 $A$ GeV. The beam impinged on a liquid hydrogen target of 87.3 mg/cm$^2$ thickness, which was enclosed in a thin titanium casing [32]. A thin aluminium beam monitor was placed in front of the target. In the target and in the beam monitor, the primary beam looses a few percent of its energy; thus corrections due to energy loss do not deteriorate the accurate measurement of the longitudinal momenta of the reaction residues. The reaction products entered into the fragment separator (FRS), used as a high-resolution spectrometer. The FRS [33] is a two-stage magnetic spectrometer, achromatic at the exit and dispersive in the central image plane. It has a momentum acceptance of 3% and an angular acceptance of 15 mrad around the beam axis. At the intermediate image plane, the fragments pass through a layer of matter (a scintillator, in our case). The energy loss of the fragments depends mostly on their charge. Every fragment will reduce its velocity and consequently its magnetic rigidity according to its atomic number. Due to the limited momentum acceptance of the FRS, only a selected number of ions, with certain atomic numbers, will have the adequate velocity to be transmitted along the second section of the FRS. This selection in Z forced us to divide the experiment in four measurements [this work, 11, 12, 13], according to the transmitted fragments in the second section of the FRS. In the present work, only fragments with atomic number around Z=20 could be transmitted. This limits the results between Z=7 and Z=37. The selection in Z turned out to be very useful for the measurement of light products. Their production cross sections are low compared to those of residues with higher mass and similar rigidity. In order not to overload the detectors, the intensity of the beam would have been limited by the high counting rate of the heavier fragments, and consequently the low counting rate of low-mass residues would have caused a large statistical error.

The essential detector equipment consisted of two scintillators, placed at the intermediate and final planes, and two ionisation chambers, placed at the exit (see Fig. 1). Multiwire detectors placed in every image plane were used for beam monitoring and calibrations, but most of them were not in the beam line during the measurements.

The scintillation detectors were used to determine the time-of-flight of the fragments and their horizontal position ($x$-position). The time-of-flight, together with the flight path, was used to deduce the velocity of the fragment. The $x$-position gave the effective radii of the trajectory, which, multiplied by the value of the magnetic field, provided the magnetic rigidity of the fragment. Full identification of the reaction residues was performed by determining the atomic number $Z$ from the energy-loss measurement with an ionisation chamber, and the mass-over-charge ratio, $A/Z$, from the magnetic rigidity and the velocity, according to the equation:

$$\frac{A}{Z} = \frac{e}{u} \frac{B\rho}{\beta_{tof}\gamma_{tof} c} \quad (1)$$

where $A$ is the mass number, $Z$ is the atomic number, $B$ is the magnetic field inside the magnet, $\rho$ is the radius of the trajectory, u is the atomic mass unit, -e is the electron charge, $\gamma_{tof} = 1/\sqrt{1-\beta_{tof}^2}$ with $\beta_{tof} = \upsilon_{tof}/c$, where $\upsilon_{tof}$ is the velocity of the ion, determined with the time-of-flight measurement, and c is the velocity of light. The energy-loss signal, $\Delta E$, was corrected for the velocity dependence of the ion and for the recombination losses. Since the nuclear charges of the fragments analyzed here are low and their velocity is high, practically all the fragments were completely stripped, and the ionic charges coincided with the nuclear charges without further corrections. The calibration of the atomic number of the products was performed exploiting the parabolic dependence $\Delta E \propto Z^2$, whose minimum is in correspondence with $Z=0$. In Fig. 2, the identification pattern for the nuclides analyzed in this work is presented. The pattern directly gave the mass calibration, thanks to the characteristic vertical line at $A/Z=2$. An additional calibration of the atomic charge, independent from the previous one, was obtained by means of the identification pattern of Fig. 2. The inverse of the length of the generic horizontal line of Fig. 2, which represents the distance among two close isotopes, is represented by the following variable, $V$:

$$V = \frac{1}{((A+1)/Z)-(A/Z)} = Z \quad (2)$$

which directly gives the atomic number $Z$.

Once the reaction residue was identified, i.e. $A$ and $Z$ are exact integer numbers with no error associated, the measurement of the magnetic rigidity $B\rho$ in the first half of the FRS, deduced from the horizontal position at the intermediate dispersive image plane, gives a precise information on its longitudinal velocity $\upsilon$ according to the equation:

$$\beta\gamma c = B\rho \cdot \frac{e}{u + \delta u} \frac{Z}{A} \quad (3)$$

where $\beta = \upsilon/c$ and $\gamma = 1/\sqrt{1-\beta^2}$. $\delta u$ is the mass excess per nucleon, which was neglected for the purpose of mass identification (Eq. (1)) but has to be taken into account here to have a more precise result for the velocities. The magnetic fields are measured by Hall probes with a relative precision of $10^{-4}$. The bending radius $\rho$ is deduced from the position of the reaction products at the intermediate image plane with a relative uncertainty of about $\pm 2 \cdot 10^{-4}$, based on a resolution of FWHM ≈ 3 mm in the measurement of the horizontal position. This results in an uncertainty of $\pm 2 \cdot 10^{-4}$ in the longitudinal momentum of individual reaction products.

Details of the experimental set-up, in particular the fragment separator and the detector equipment [34,35] as well as a description of the analysis method [36,37,6] can be found in previous publications. Details of the experimental procedure and of the data analysis technique used in this work are documented in the underlying thesis [38].

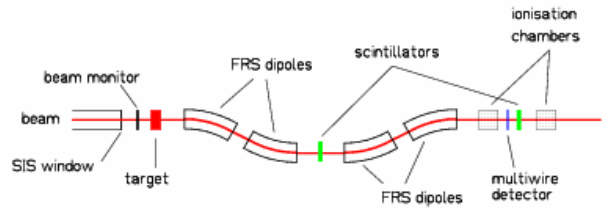

Fig. 1: Schematic view from above of the horizontal section of the experimental set-up. In the analysis, the orthogonal Cartesian reference axes were set as follows: z-axis along the beam line, y-axis perpendicular to the sheet and x-axis on the plane of the sheet.

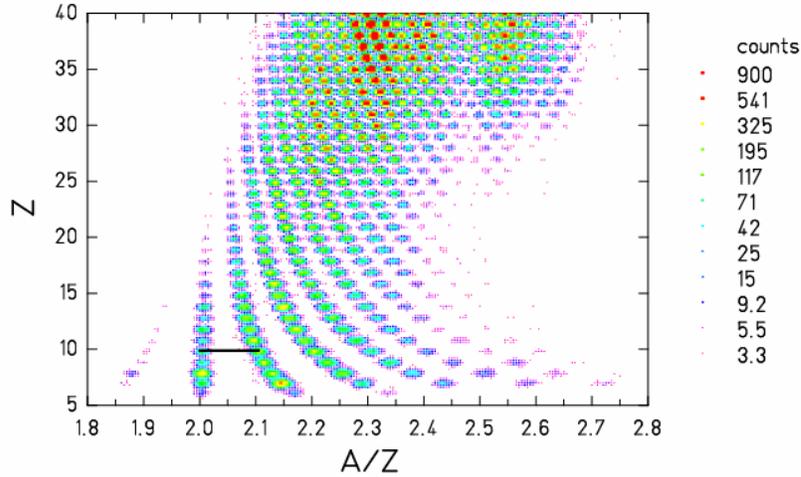

Fig. 2: (Colour on-line) Identification pattern of the measured data. The vertical line at *A/Z*=2 gives the mass calibration. The inverse of the length of the generic horizontal line, which represents the distance among two close isotopes, gives the atomic number *Z* (10 in the example) and provides the charge calibration. The plot collects the counts, given in logarithmic scale, from the hydrogen target, including the contribution from the titanium windows of the container, for the present measurement (fragments with atomic number around Z=20).

### 3. Data analysis

Considering the limited momentum acceptance of the FRS of 3%, only part of the velocity spectra of a restricted number of nuclides was measured at once. In order to fully cover the momentum distributions of all residues, measurements obtained by changing the magnetic fields in steps of 3% were combined. To do this, the spectra were normalised before to the number of beam-monitor counts and corrected for dead-time losses of the data acquisition. With this procedure, the longitudinal-velocity spectra were obtained for every observed nuclide. In Fig. 3-left, the velocity distribution of $^{58}$Fe in the beam frame is presented as an example. Combining the longitudinal-velocity spectra of all isotopes of one element, the two-dimensional cluster-plot of the velocity distribution as a function of the neutron number was obtained for every element. In Fig. 3-right, the cluster-plot velocity distribution of iron is presented as an example. One may notice that the counts are grouped into 8 different transversal bands, as for instance that one enclosed inside the dashed line, corresponding to 8 different $B\rho$ measurements. The missing band would contain products with the same magnetic rigidity as the beam, which could not be measured. The spectra in Fig. 3 include the contribution from the titanium windows of the container. An additional experiment, performed using a titanium target of the same thickness as the windows of the liquid-hydrogen container, provided the background production, which was subtracted to obtain the yield in hydrogen. From the data taken with the titanium target we deduced that the nuclei with the most extreme velocity values (covering the external wings of the velocity spectra in Fig. 3-left) are mostly due to the interaction of uranium with protons, while the central part is exclusively due to the interaction of uranium with titanium.

The spectra of Fig. 3 must be observed keeping in mind that, due to the limited angular acceptance of the FRS (15 mrad around 0º), represented by a cone in the laboratory frame, only the part of the production inside the cone is transmitted through the FRS and can actually be observed. According to what was found in previous experiments for similar systems [5, 6, 7, 22, 37, 39, 40, 41], we assume that the situation can schematically be described as depicted in Fig. 4.

The three humps of the velocity distributions were interpreted as fission fragments emitted in backward direction (left peak), fragmentation products (central peak) and fission fragments emitted in forward direction (right peak). The interpretation was justified by the following considerations. The velocity distribution of the fragmentation residues is represented in the beam frame by a three-dimensional Gaussian [42]. A Gaussian-like shape is the result of the statistical superposition of several momentum contributions in space, attributed to the momenta of abraded nucleons [43] and to the recoil of evaporated particles. In peripheral collisions, due to the abrasion, the longitudinal mean value is expected to be slightly negative with respect to the beam velocity [44]. When the fragment is produced in a fission event, the kinetic energy that it acquires is more or less fixed, assuming that the fissioning nuclei belong to a limited range in $Z$ and $A$, so the possible values of its velocity cover only the external shell of a sphere. The centre of the sphere represents the mean velocity of the fissioning nucleus. In the beam frame it is slightly negative, because of the preceding abrasion or intra-nuclear-cascade process. The radius of this sphere results from the Coulomb repulsion between the fission fragments and momentum conservation, and thus it provides information on the mass and charge of the complementary fragment. this scenario also explains that the peak at positive velocities, corresponding to forward-emitted fission products, is higher than the peak of backward-emitted fragments, due to the larger transmission of the FRS. Please note that a similar pattern in velocity space can result also from other kinds of reactions like evaporation of light fragments or break-up reactions with one heavy remnant. Keeping this remark in mind, in the following, for simplicity, we will call the reaction products showing this kind of kinematical pattern "fission fragments", but we will come back to a more general discussion later.

The interpretation of the data as fission and fragmentation products is consistent also with the characteristics of the isotopic distributions that could be extrapolated from Fig. 3-right: the two peripheral humps (fission) are shifted to the right with respect to the central hump (fragmentation). As expected, fragmentation generally produces nuclei on the neutron-deficient side of the beta-stability valley, while in fission processes more neutron-rich fragments are produced. The velocity spectra observed inside the limited angular acceptance of the FRS turn to be a useful tool to disentangle the different reaction mechanisms.

The shape of the velocity spectrum of every element was reconstructed by overlapping all the velocity distributions of the isotopes of that element. The overlapping was done comparing channel by channel all the velocity distributions of the isotopes and taking the maximum value. This corresponds to overlap the distributions of all the isotopes and draw the skyline. In this way, to the velocity distribution of every element contribute its most-produced fragmentation isotope and its most-produced isotope by fission processes. From a graphical point of view, this procedure corresponds to squeeze all the isotopes of the two-dimensional cluster plots of the velocity distributions, like that one of Fig. 3-right, in one line. Combining the spectra of all the elements together, the two-dimensional cluster-plot of the velocity distribution as a function of the produced elements could be constructed (Fig. 5). Fig. 5 includes the contribution of the titanium windows, which is responsible for fragmentation products filling the central band of the distribution. The unexpected trend of the mean velocity of the fragmentation products, which increases with decreasing mass, was discussed in a separate publication [45].

The data analysis was based on the reconstruction of the full velocity distribution of each isotope. The extrapolation of the quantitative information from the raw spectra was not possible because some data were missing, as visible in Fig. 3-right. A fit with three Gaussian curves was used to reconstruct the full spectrum for every isotope (see Fig. 3-left). The procedure was optimised by fitting the data of every isotope on the base of the results of a common fit obtained by fitting of all data at once, as explained in ref. [38]. From the result of the fits, the mean values and the standard deviations of the longitudinal velocity spectra and the yields inside the angular acceptance of the FRS could be determined for the fragmentation products and for fission fragments emitted in backward and in forward direction.

The yields for the $^{238}$U+$^{1}$H system were deduced subtracting the background yields obtained with the titanium dummy target. Assuming angular isotropy of the products, using the method described in ref. [46], the fraction of transmitted reaction residues can be calculated. Knowing the beam intensity, the target properties and the transmission ratios, the production cross sections were calculated on the basis of the measured yields. An additional correction for the beam attenuation in the target was considered in the evaluation of the cross sections. The effect of secondary reactions in the target was in most cases negligible; however the error bars were increased in order to account for this uncertainty. The contribution of secondary reactions was determined as described in the appendix of ref. [6].

The finite angular acceptance of the spectrometer introduces a small deviation of the mean velocities and standard deviations from the true values. Assuming isotropic velocity distributions [9], these deviations were corrected using the transmission ratios. The mean velocity values were corrected also for the mean energy loss of the projectiles in the first half of the target and for the mean energy loss of the reaction products in the second half of the target. The average of the mean positions of the two peaks gives the mean recoil velocity of the mother nucleus in the beam frame introduced in the nuclear reaction. The mean value of the velocity of the fission fragments in the frame of the fissioning nuclei corresponds to the absolute value of the difference between any mean positions of the two velocity peaks and the mean recoil velocity.

Below $Z=17$ the forward peak of the double-humped distribution could not be unambiguously determined due to the contribution from the titanium target in this range. Therefore, the mean values of the velocity of the fission fragments are given only in the range $17 \leq Z \leq 39$. The backward hump was observable without discontinuity for $7 \leq Z \leq 39$. This fact permitted to obtain the fission cross sections for the entire $Z$ range, using the yield obtained from the fit of the backward hump. However, the results for $Z=38$ and $Z=39$ were excluded because they were at the extreme of the $B\rho$ selection, falling at the border of the scintillator at the exit of the FRS, with the consequence that part of the production could not be detected and their cross sections are systematically underestimated.

With the above described technique, the cross sections and the velocity distributions were measured for both reaction mechanisms (fission and fragmentation) for every nuclide produced in both systems ($^{238}$U+$^{1}$H and $^{238}$U+Ti).

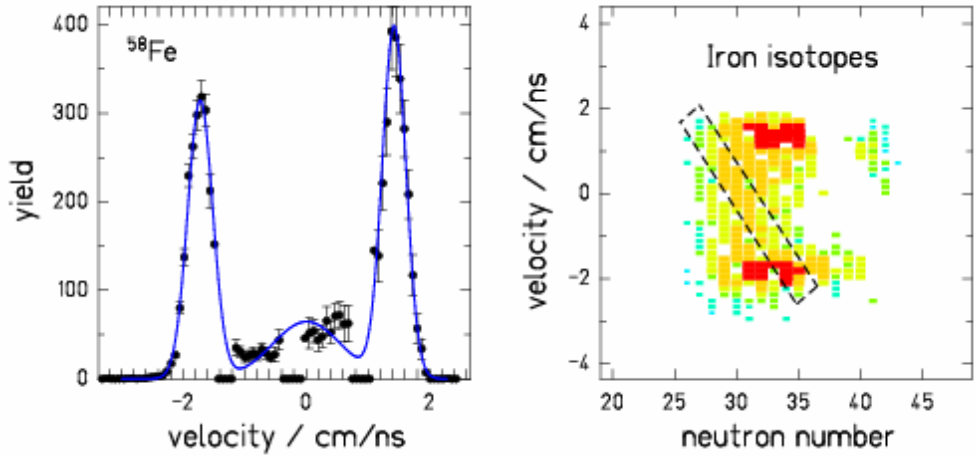

Fig. 3: (Colour on-line) Left: Longitudinal-velocity spectrum for $^{58}$Fe. The experimental data were fitted with three Gaussian curves. Right: Two-dimensional cluster-plot of the velocity distribution as a function of the neutron number for iron ($Z=26$). The dashed line encloses the events collected in one $B\rho$ measurement. In both figures, the data refer to the interaction of the uranium beam with $^{1}$H + Ti. The velocity is presented in the beam frame ($\upsilon_{238U} = 0$ cm/ns). The counts are normalised to the beam dose.

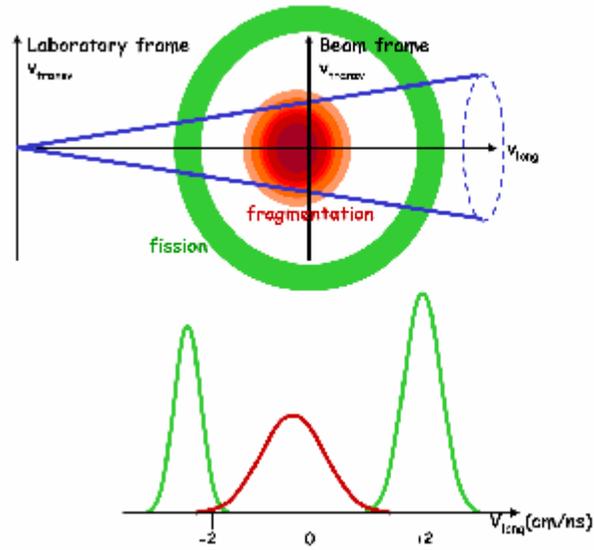

Fig. 4: (Colour on-line) Up: schematic representation of the velocity distributions of fragmentation and fission residues of one isotope, together with the FRS angular acceptance. Down: projection of the accepted events on the longitudinal axis.

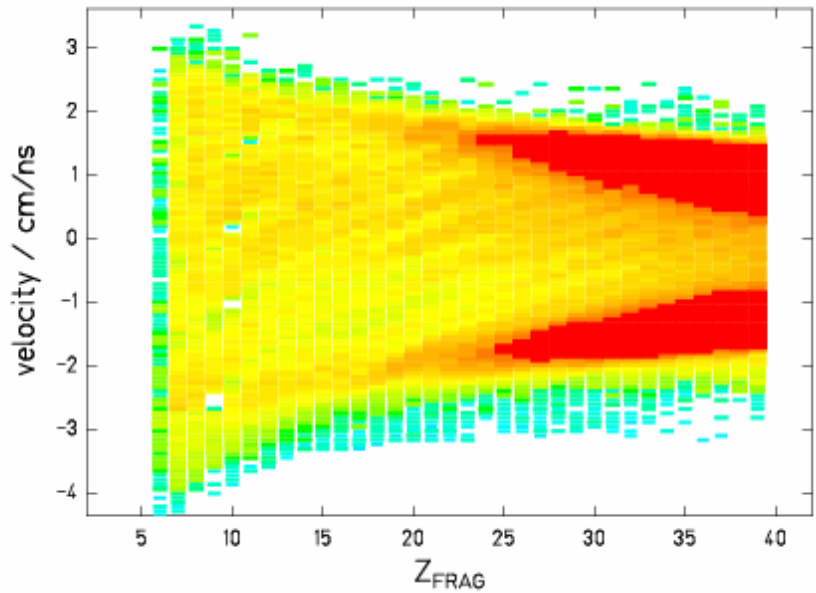

Fig. 5: (Colour on-line) Two-dimensional cluster plot of the experimental velocities of fragments produced in the interaction of the uranium beam with the hydrogen-plus-titanium-window target. The velocity is presented in the beam frame ($v_{238U} = 0$ cm/ns).

## 4. Results

### 4.1 Measured production cross sections

In Fig. 6, the measured cross sections are presented in form of isotopic distributions in the range $7 \leq Z \leq 37$. The numerical results are collected in Table A.1 of the appendix. The statistical error was determined by the error associated to the fitted parameter in the individual fit of the velocity distribution of every nuclide. In turn, the latter reflects the statistical uncertainties associated to the single data points forming the velocity distribution, which were determined by the inverse of the square root of the number of counts, according to the Poisson statistics. In order to account for the eventual deviations of the fit function from the "true" shape of the velocity distribution, the statistical uncertainty was increased until the square root of $\chi^2$ divided by the degrees of freedom was approximately 1. Due to the asymmetry of the Poisson distribution, a confidence interval of 68% is not symmetric around the most probable value. This fact results in asymmetric error bars. This asymmetry is important for low counting and tends to disappear in case of a large number of events. The systematic uncertainties are due to: the uncertainty on the width and mean value of the velocity distributions, the calibration factor that converts the beam-monitor counts into the number of $^{238}$U projectiles, the evaluation of the angular transmission, the thickness of the target, and the secondary reactions. The production cross sections of some nuclides are missing, because the magnetic-field settings for those nuclei were not performed. In addition, some data had to be discarded for technical reasons. Those isotopes, for which the contribution of secondary reactions was estimated to be high, were also discharged.

The dashed lines in Fig. 6 were obtained with an interpolation by smoothing the existing data. When a data point is missing, the dashed line has to be taken just as a guideline for the eye. The dashed lines may not represent the real physical content, since the data do not necessarily follow a smooth behaviour, as explained in ref. [47]. In Fig. 7, the cross sections for all the nuclides analysed in this work are presented on the chart of the nuclides. As explained in the introduction, in the frame of the same experiment, three other works proceeded in parallel to analyse the data: in the fission region ($28 \leq Z \leq 64$, [12], and $65 \leq Z \leq 73$, [13]), and in the fragmentation region ($74 \leq Z \leq 92$, [11]). In this systematic study, the present work covers the part of the lightest nuclei ($7 \leq Z \leq 37$). In the region where two measurements overlapped (from $Z=28$ to $Z=37$) the experimental results generally agree within the error bars. The first, almost complete, general presentation on the preliminary data was discussed in ref. [48]. The complete overview on residual-nuclide production cross sections is presented in ref. [20] as a cluster plot on the chart of the nuclides. The numerical values for the entire set of data are available in ref. [49]. It represents the most complete residual-nuclide distribution of a proton-induced spallation reaction on uranium ever obtained.

Regarding the region of light masses, three interesting aspects can be noticed from Fig. 6 and from the overview of Fig. 7. Firstly, the isotopic distributions are long and shifted towards the neutron-rich side for the heavier fragments; they shorten and move towards stability as the mass decreases. As a second interesting fact, we observed that the production extends down to very light fragments. Our measurement was technically limited to $Z \geq 7$, but the production seems to extend even farther down. A third feature is the height of the cross sections. As expected, the cross sections are very high in the main fission region and decrease rapidly from $Z=30$ to $Z=20$. But then they stay constant and finally slightly increase again below $Z=10$. A discussion on these features will be presented in section 6.2.

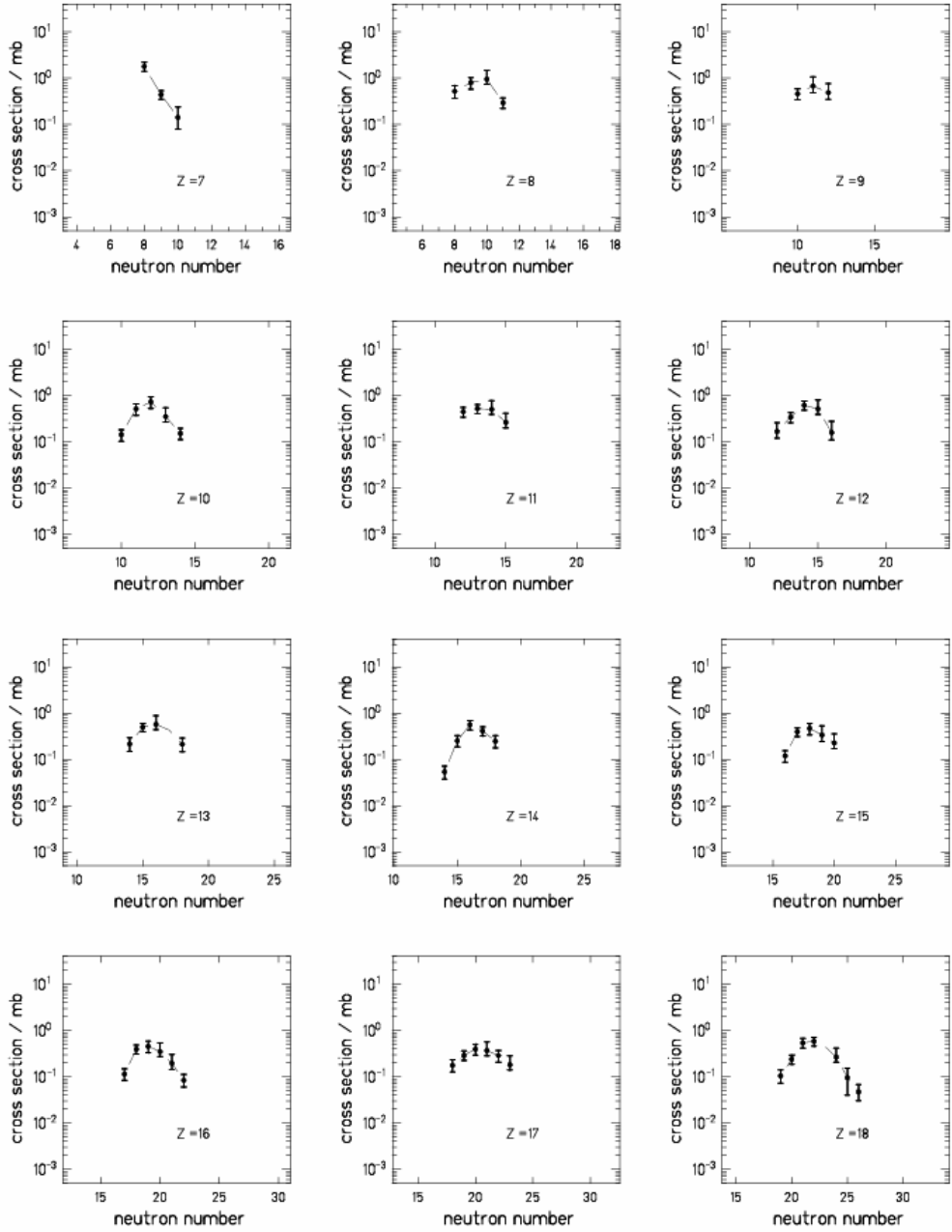

Fig. 6: Isotopic cross sections for the products between $Z$=7 and $Z$=37 produced in the reaction $^{238}$U on proton at 1 GeV per nucleon. The dashed lines are set to guide the eye and do not necessarily represents the expected trend of the missing data. The error bars include both the statistical and systematic uncertainties.

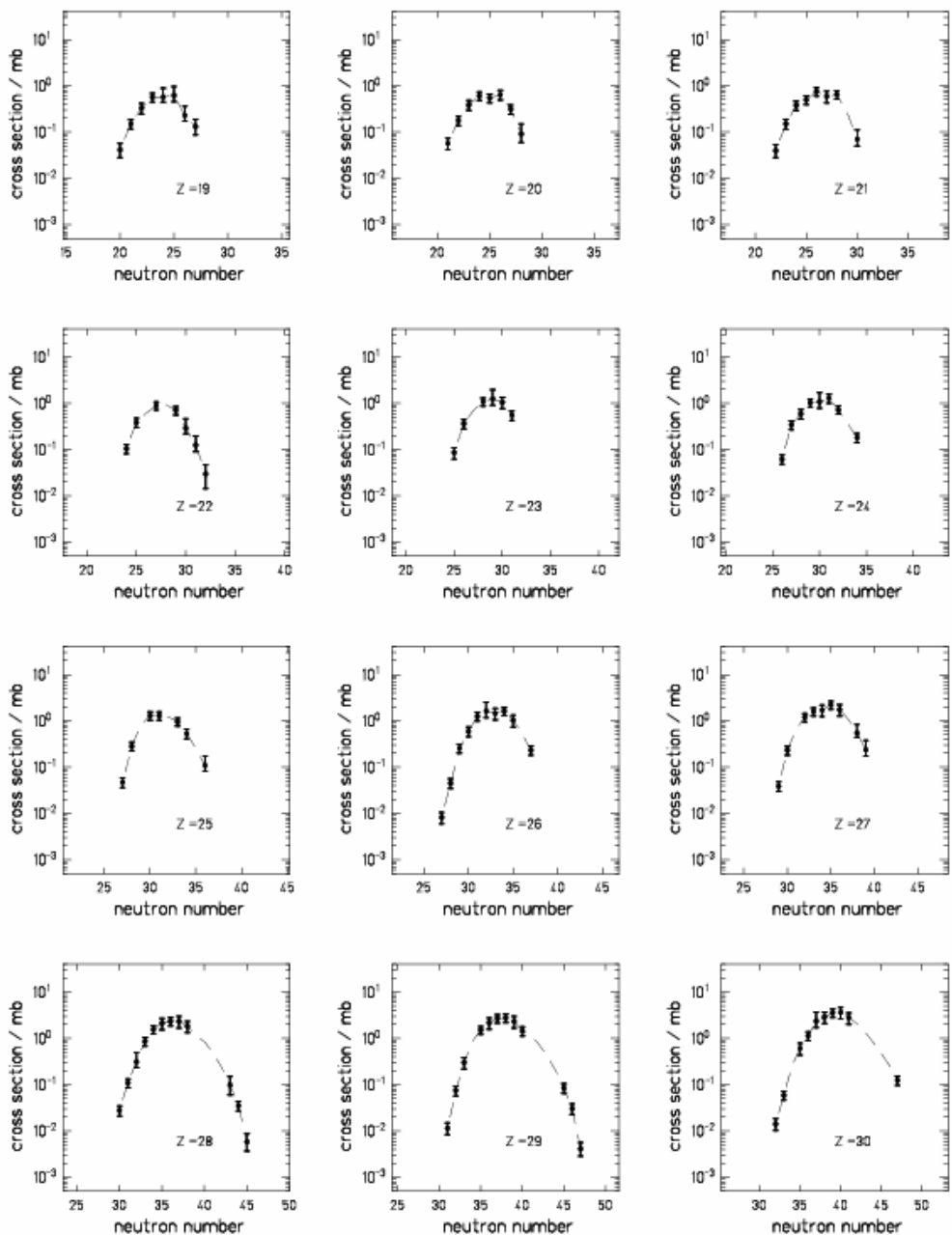

Fig. 6 (continue)

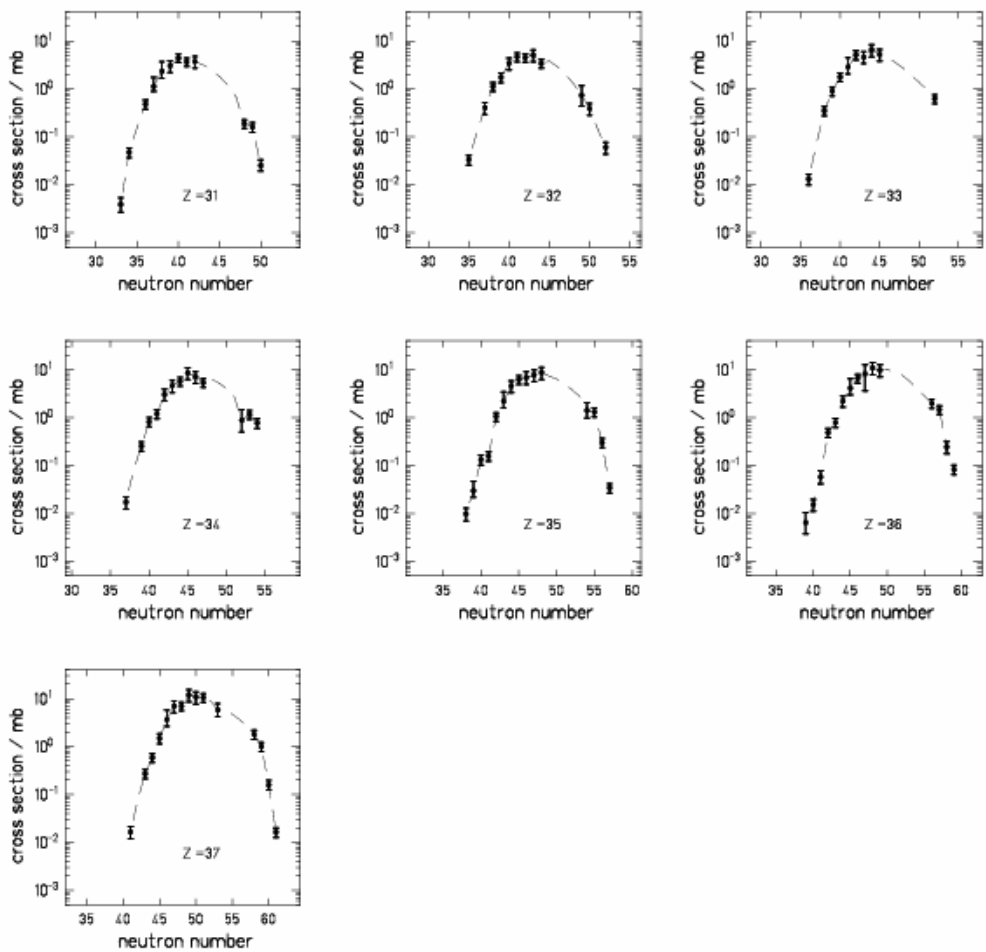

Fig. 6 (continue)

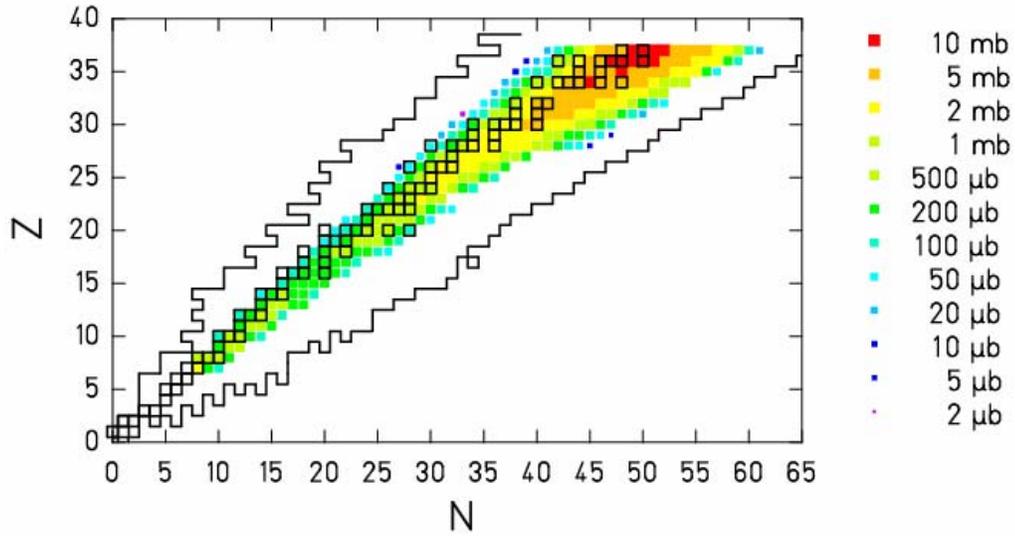

Fig. 7: (Colour on-line) Two-dimensional cluster plot of the nuclide production cross sections in the reaction $^{238}$U + $^{1}$H at 1 $A$ GeV, obtained in this work, on a chart of the nuclides. The numerical values of the measured data are collected in Table A.1. For those nuclides, which could not be measured, the cross sections were interpolated from the existing data by smoothing the isotopic distributions. The open black squares correspond to stable nuclides. The lines indicate the limit of the known nuclides.

*4.2 Velocity distributions*

The velocity distributions of the fragments contain other valuable information on the nuclear-reaction aspects. Although small changes of the mean velocities and of the standard deviations in the isotopic chain of one element are expected, the results will be presented as a function of the atomic number of the nuclides because no variation was observed inside the error bars among isotopes of the same element. The mean velocity of the fragments, presented in the frame of the mother nucleus, is shown in Fig. 8-down. The mean recoil velocity of the mean mother nucleus in the beam frame is shown in Fig. 8-up. Both figures include also the data obtained in the complementary analysis centred around symmetric fission [12]. As already said, the data stop at $Z$=17 because the forward peak was not clearly disentangled below $Z$=17. In both figures, the error bars which are not visible are smaller than the data points. Numerical values are collected in Table A2 of the appendix. Table A2 collects also the numerical values of the kinetic energies deduced from the data of Fig. 8.

In Fig. 9, we present the standard deviations of the two peaks of the velocity distributions of the fragments observed in forward and backward directions for $Z \geq 17$. The widths of the two peaks of the velocity distributions of the lighter nuclei in backward direction with $Z < 17$ could not be determined with good precision, mostly due to the relatively large correction for the production in the titanium windows. These data do not give direct information on the physics of the reaction, since the widths of these distributions are affected by two main disturbing contributions. One is the finite angular range accepted by the FRS, which introduces an increase in width in the longitudinal momentum (see Fig. 4). This contribution is larger for higher nuclear charges [46], more transmitted than the lower ones. The difference in energy loss of projectile and fragments in the target before and after the reaction introduces another energy broadening of the residues, named "location straggling" [50], which slightly decreases with increasing nuclear charge. Both effects depend dominantly on the atomic number, $Z$. In Table 1, the measured widths of the backward velocity humps have been corrected for these two contributions for two nuclei, Ar and Sr (in the calculation two isotopes were used). The energy loss was calculated with the program AMADEUS [51] and the effect of transmission was estimated as explained in ref. [46]. For a specific nuclide, the relative width in velocity induced in the reaction ($\sigma_v^{react}/v$) results to about 9-10 %, approximately constant over the whole range of elements. This corresponds to a relative width in kinetic energy of the fragments of about 18-20 %. For the heavier fission products ($Z \geq 30$), the absolute velocity width $\sigma_v^{react}$ remains constant at about 0.125 cm/ns [12].

The values of $\sigma_v^{react}$ include three contributions which cannot be disentangled from the present experiment. The first one emerges from the variation of the total kinetic energy (TKE) for a given fissioning system. The second one is caused by the different fissioning systems contributing to the production of a certain fission fragment. The third one is caused by the fluctuations of the velocity of the prefragment due to the Fermi momenta of the removed nucleons. These latter aspects will be discussed also in section 6.3.

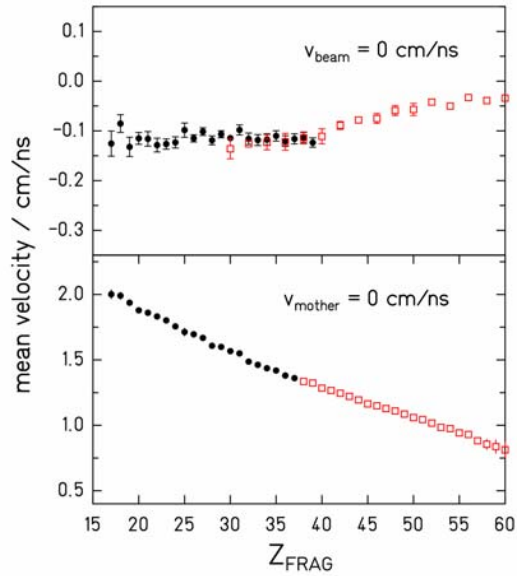

Fig. 8: Experimental results for the reaction $^{238}$U + $^{1}$H at 1 $A$ GeV. Up: Mean recoil velocities of the mother nuclei of all fragments measured in this experiment presented in the beam frame: this work (full dots), ref. [12] (empty squares). Down: Mean values of the velocities of the fragments in the frame of the mother nuclei: this work (full dots), ref. [12] (empty squares). Values are drawn as a function of the atomic number of the fragment.

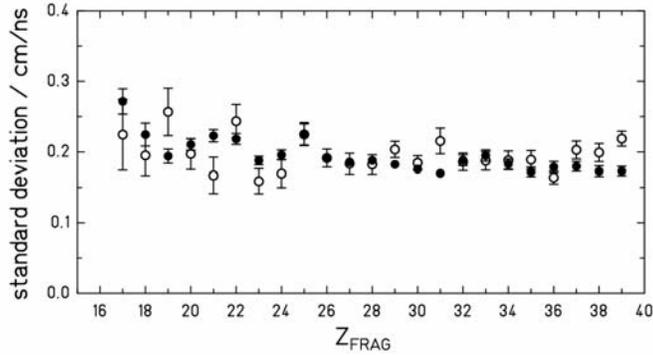

Fig. 9: Measured standard deviations, $\sigma_\upsilon^{meas}$, of the velocities of fragments emitted in backward (full dots) and forward (empty dots) direction, produced in the reaction $^{238}$U + $^{1}$H at 1 $A$ GeV. The lines are the results given by the fitting procedure. The data are affected by the FRS angular transmission and by the location straggling in the target (see text). Values are drawn as a function of the atomic number of the fragment.

Table 1: Contributions to the measured width $\sigma_\upsilon^{meas}$ of the backward peak of the velocity distribution of two fragments: due to location straggling ($\sigma_\upsilon^{\Delta E}$) and due to the variation of the longitudinal velocity in the transmitted angular range ($\sigma_\upsilon^T$). The velocity width caused by the nuclear reaction $\sigma_\upsilon^{reac}$ is deduced from the quadratic subtraction of the two terms. Combining the results for $\sigma_\upsilon^{reac}$ with those of Tab. A.2 one obtains the relative velocity width and the relative kinetic energy width (last two columns).

|  | $\sigma_\upsilon^{meas}$ | $\sigma_\upsilon^{\Delta E}$ | $\sigma_\upsilon^T$ | $\sigma_\upsilon^{reac}$ | $\dfrac{\sigma_\upsilon^{reac}}{\upsilon}$ | $\dfrac{\sigma_E^{reac}}{E}$ |
|---|---|---|---|---|---|---|
| $^{40}$Ar ($Z$=18) | 0.21 cm/ns | 0.062 cm/ns | 0.092 cm/ns | 0.18 cm/ns | 9 % | 18 % |
| $^{90}$Sr ($Z$=38) | 0.19 cm/ns | 0.045 cm/ns | 0.118 cm/ns | 0.14 cm/ns | 10 % | 20 % |

## 5. Comparison with other data

### 5.1 Nuclide production

Data on the production of nuclides by nuclear fission, fully identified in $Z$ and $A$, are scarce. Before the use of inverse kinematics, the measurement of the formation cross-sections of individual nuclides mostly relied on their radiochemical properties and on the online mass-separator technique. In most counter experiments only mass distributions are obtained. In a recent experiment with secondary beams a large number of element distributions has been determined [52], however no mass identification could be given. Only a few experiments on thermal-neutron-induced fission, performed at ILL, Grenoble, have given a rather comprehensive overview on the nuclide production in the light fission-fragment group for a few odd-N fissile systems [53], however not extending below $Z \approx 26$ [54]. Data of excellent quality on nuclide production from higher excitation energies only exist for fission induced by relativistic $^{238}$U projectiles in various targets, e.g. [37,41], but they did not extend to very light elements.

For the above reasons, there are very few experimental cross sections available, forming a full isotopic distribution, comparable with our data. One of these few is that one of rubidium ($Z$=37), measured by Belyaev et al. [55] in 1980. These data have already been compared with the results of the present experiment in ref. [12], showing a good agreement. Yields of very light nuclides produced in interactions of 600 MeV protons with $^{238}$U were already observed in direct kinematics [56]. As an example, in Fig. 10 the distribution of the potassium isotopes obtained in our experiment is compared to the yields measured at ISOLDE from 600 MeV protons in a thick uranium-carbide target [57]. The yields from the ISOLDE experiment (scale on the right) were normalised to our cross sections (scale on the left). The difference in energy is not expected to produce a significant difference in the cross sections [58]. The isotopic distribution is quite neutron rich with respect to the valley of beta-stability. Since the ISOL method provides high efficiencies for a limited number of elements only, there was no knowledge on the overall nuclide production in the target from these measurements. Fig. 7 can be considered a sort of "map" of the potentially available light radioactive beams by proton-induced reactions using a $^{238}$U target. In addition, the systematic results offered by our measurements can be exploited for the determination of the efficiency of the ISOLDE technique.

The data of ref. [57] were measured in direct kinematics. The experiments could not supply any information on the velocities, thus there was no knowledge on the reaction process that produced them. The velocity characteristics of the data measured in the present experiment indicate that the potassium isotopes presented in Fig. 6 formed in proton-induced spallation of $^{238}$U at 1 $A$ GeV originate from the binary decay of a heavy nucleus. We can deduce that also the data of ref. [57] have the same kinematical characteristics.

In 1958 the production of $^{24}$Na from proton-induced reactions on several targets at several energies was investigated [59]. The result for 1 GeV protons on $^{238}$U can be compared with our data. The two measurements give: (0.63±0.16) mb [59] and (0.53±0.12) mb [this work]. The results agree within the error bars.

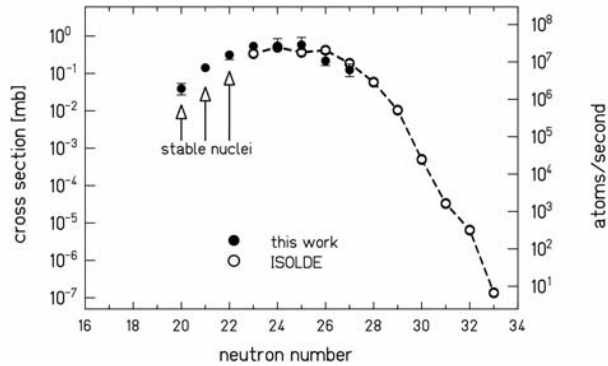

Fig. 10: Measured cross sections of potassium isotopes ($Z$=19) from 1 $A$ GeV $^{238}$U in hydrogen of this work (full dots) are compared with the yields of potassium isotopes from the reaction of 600 MeV protons in a thick uranium-carbide target (open dots), measured at ISOLDE [57]. The yields from the ISOLDE experiment (right scale) were normalised to the cross sections of this work (left scale).

*5.2 Mass and charge distributions*

A few additional rather dispersed data are available for yields of nuclides from reactions of protons with heavy nuclei. Most of them were measured with radiochemical detection methods in experiments performed in direct kinematics [60, 61, 62, 63, 64, 65]. Only in few cases, as for instance for 340 MeV protons on tantalum [66], the mass distribution, deduced from the experimental data, extended with continuity from the heavy to the very light fragments, forming a W-shaped distribution (see figure 6 of ref. [66]). In an experiment performed at LEAR (Low Energy Antiproton Ring) at CERN [67], the mass distribution of fragments produced in the antiproton-induced fission of $^{238}$U nuclei was obtained. The fission products could be selected by their kinetic energy and by a coincidence condition. The mass spectrum shows a minimum between $A$~20 and $A$~40 (see figure 2 of ref. [67]). A similar behaviour was observed also in the binary decay of a $^{244}$Cm compound nucleus [68] produced in the heavy-ion fusion reaction of 8.4·$A$ MeV $^{232}$Th on $^{12}$C (see figure 3 of ref. [68]). The mass and charge distributions of the binary-decay products observed in this work also present a similar shape, as can be seen in Fig. 11, where the data are presented together with the other fission products measured in this experiment, analysed in a separate work [12]. These results will be discussed in section 6.

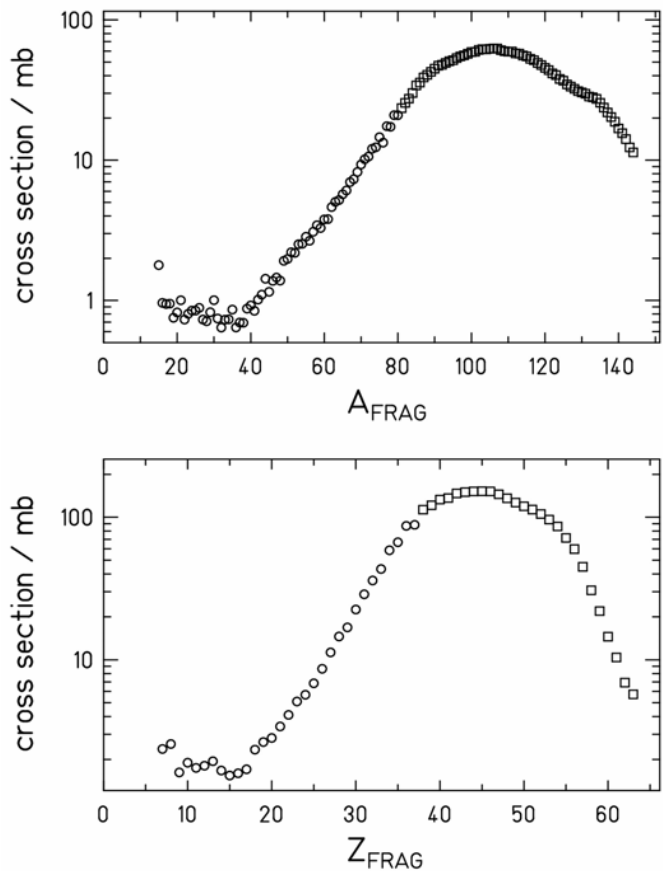

Fig. 11: Mass and charge distributions of binary-decay products measured in the reaction of $^{238}$U+p at 1 GeV. Dots: this work; squares: taken from ref. [12] in the range not covered by the present work.

*5.3 Velocities*

In the review "Fission of highly excited nuclei", Andronenko et al. [65] collected a large amount of experimental data for reactions induced by 1 GeV protons. Among other results, the review summarizes data on angular distributions, mean longitudinal momenta and kinetic energies. Besides the review of Andronenko such kind of data are reported and discussed in several publications, e.g. [25], [69], [70], and [31]. However, we could not find any measured data directly comparable with the light fragments produced in $^{238}$U + $^{1}$H at 1 *A* GeV, analysed in this work.

**6. Discussion**

*6.1 Possible reaction mechanism*

In this section, the reaction mechanisms behind the production of light nuclides in spallation reactions will be discussed. In this context, it is useful to have an overview of the whole production, presented on the chart of the nuclides in Fig. 7 and in ref. [20].

The proton-rich heavy evaporation residues, filling the upper part of the chart of the nuclides, are kinematically characterised by narrow, Gaussian velocity distributions, with mean values close to the velocity of the projectile. One may wonder whether the light fragments observed in the present work could be spallation-evaporation residues. Selecting kinematically the evaporation residues, Taïeb et al. [11] proved that the spallation-evaporation corridor in $^{238}$U + $^{1}$H at 1 *A* GeV dies out rather soon (around *Z*=74).

As already discussed, due to the angular cut of the FRS, the fission fragments are characterised by a double-humped distribution of the longitudinal velocity. The shorter is the distance between the two humps the smaller is the velocity of the fragment [12]. This distance decreases as the charge of the fragment increases. At *Z*>63, the two humps start to merge and form a single hump. Up to the last element investigated (*Z*=74) the width of this hump is too large to be interpreted as the velocity spectra of just a spallation-evaporation residue. Thus the large width of the hump indicates a fission fragment and is consistent with the existence of a complementary light fission partner with large velocity, the fragments observed in the present work.

In a consistent way with what was observed in [12], the present experiment proves by the velocity distributions that the light nuclides in the spallation of $^{238}$U by 1 GeV protons are produced together with a complementary heavy residue. Taking into account the conservation of the momentum, the Coulomb repulsion between the light nuclide and its complementary heavy residue explains the large velocity in the beam frame. These velocities follow on with continuity the pattern indicated by the heavier fission fragments (Fig. 8-down). Also the charge and mass distributions of Fig. 11 do not present any discontinuity. All the experimental evidences indicate that the light residues observed in this work are fission fragments. As it will be discussed in the next section, it is even expected from theoretical considerations and proven by several experiments (e.g. [67, 68]) that mass distributions in fission show increased production yields for very asymmetric mass splits. These very asymmetric splits have been interpreted as a natural transition from fission to evaporation.

The fission-evaporation mechanism is for sure responsible for the production of light fragments. However, one may question whether this is the *dominant* production process. In the past, it was discussed [31] if the production of the light nuclides could be due mostly to a fast binary decay right after the intra-nuclear cascade phase, before a fully thermalised compound nucleus is formed. Such a process would release residual nuclei having similar characteristics as the fragments observed in the present work: large velocities − increasing as the mass decreases − and rapidly increasing cross sections with decreasing *Z*, below *Z*=10.

In view of these considerations, we will discuss at first the role of fission-evaporation in the production of the light residues (section 6.2). At the end, the contribution of a possible break-up channel will be discussed (section 6.3).

*6.2 Fission*

*6.2.1 Transition from fission to evaporation*

According to the transition-state model, the decay rate for fission depends on the properties of the fissioning nucleus in the "transition state", i.e. on the phase space available in the saddle-point configuration. The saddle point represents a kind of bottleneck through which the nucleus is forced to pass on the way to fission [71]. At the saddle point the potential energy, $U$, associated with the shape (deformation, $\varepsilon$) of the nucleus, $U(\varepsilon)$, has reached a maximum. The height in energy of this maximum with respect to the ground state of the nucleus is the fission barrier, $B_{fiss}$. The potential energy depends also on mass-asymmetric deformations, which lead to the formation of two fragments of different sizes [72]. The relation between mass-asymmetry deformation at saddle and mass split at scission is assumed to be essentially strict and undisturbed by fluctuations due to the dynamics of the system between saddle and scission [73]. If $A_1$ and $A_2$ are the masses of the two fragments, the mass-asymmetric deformation can be expressed in terms of the "mass asymmetry parameter", $\eta = A_1/(A_1+A_2)$. Consequently, also the fission barrier can be calculated for every mass asymmetry: $B_{fiss}(\eta)$. The potential energy forms a ridge line along the mass-asymmetry coordinate whose points are called "conditional saddle points", because of the constraint of a fixed mass asymmetry [73]. The energy of the conditional saddle points as a function of the mass asymmetry is illustratively presented in Fig. 12 for some nuclear systems. A description of the correlation of the final mass distribution and the variation of the height of the conditional saddle with mass asymmetry can be found in ref. [74].

In the statistical model of fission [75, 76], for a given excitation energy the yield of a certain fission fragment is calculated by the statistical weight of the transition states above the conditional potential barrier. This weight is in turn correlated to the density of nuclear levels. In the thermodynamic Fermi-gas picture, i.e. assuming the nucleus as a system of non-interacting fermionic particles, the density of states is in good approximation:

$$\rho(E^*_{saddle}) \propto e^{2\sqrt{aE^*_{saddle}}} \quad \text{with} \quad E^*_{saddle} = E^*_{gs} - B_{fiss}(\eta) \quad \text{and} \quad a \cong A/10 \ MeV^{-1} \qquad (4)$$

where $E^*_{gs}$ is the thermal excitation energy above the ground state and $E^*_{saddle}$ is the energy above the saddle point. The yields as a function of mass asymmetry are then given by:

$$Y(\eta, E^*_{saddle}) \propto e^{2\sqrt{a(E^*_{gs} - B_{fiss}(\eta))}} \qquad (5)$$

The result, for a heavy fissile nucleus at high excitation energies, is essentially a "W-shaped" distribution (see Fig. 12), whose maximum is at the symmetric split.

The central part of the M-shaped potential can in first approximation be described by a parabola, whose curvature, $C_\eta$, affects the width of the central part of the mass distribution, which becomes a Gaussian function, with the variance:

$$\sigma^2(\eta, E^*_{saddle/\eta=1/2}) = \frac{1}{2}\sqrt{\frac{E^*_{saddle}}{a}} \cdot \frac{1}{C_{\eta=1/2}} = \frac{1}{2}\frac{T}{C_{\eta=1/2}} \quad \text{being} \quad T = \sqrt{\frac{E^*_{saddle}}{a}} \qquad (6)$$

The excitation energy introduced in the system and the curvature of the potential affect the width of the mass distribution. Therefore, in a heavy system, the difference in intensity from the top of the yield at symmetry (for $\eta=0.5$ in Fig. 12-right), to the minimum (for $\eta=0.08$ in Fig. 12-right), is very large at low excitation energies. The consequence of this fact is that in most of the experimental observations available in literature fission seems to die out for atomic numbers below $Z \approx 28$. This is one of the reasons why the very light products (from $A=1$ to $A \sim 20$) produced in high-energy nuclear reactions have been previously attributed to a kind of fragmentation process. For a long time, fission and evaporation were treated as separate processes. Moretto [76, 77] pointed out and discussed the inconsistency of the two separate pictures and proposed that evaporation and fission should be treated as two manifestations of the same kind of binary decay with a continuous transition looking at fission in a generalised sense [78, 79].

In the review "Fission of highly excited nuclei" by Andronenko et al. [65], the mass distributions show the characteristics expected from the general properties of fission barriers as a function of mass asymmetry [72] illustrated by Fig. 12: While for heavy fissioning systems at high excitation energies symmetric fission distributions are observed, characterized by Gaussian distributions which are centred around half the mass of the mother nuclei, lighter systems show flat or even U-shaped distributions. Thus, these findings are compatible with a generalised fission process, according to the proposition of Moretto. Also the mass distributions of ref. [67] and ref. [68] were attributed to high-energy fission, extending to very large mass asymmetry. While in these two cases it was possible to verify the binary nature of the decay, for the mass distribution of ref. [66] no information on the kinematics was possible and no interpretation was proposed at that time.

To conclude, in the decay of any excited fissile compound nucleus, the full mass range is expected to be populated by binary decay, understood as a generalisation of evaporation and fission. Therefore, this process for sure contributes to the production of light residues in the spallation reaction analysed in this work. Whether it is the dominant production mechanisms or not will be discussed in section 6.3.

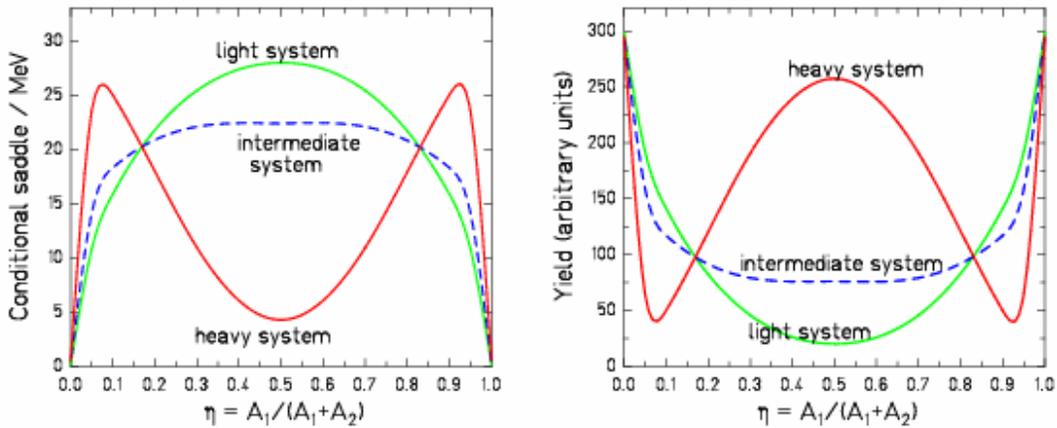

Fig. 12: Left: Schematic presentation of the fission-barrier height for a given mass split for a light, an intermediate and a heavy system. Right: corresponding yields (in arbitrary units).

*6.2.2 The mean velocities*

The quantitative reproduction of the mean velocity of the fission products is not an easy task, because several effects can affect the experimental results, as for instance the eventual presence of a third fragment, or dynamical effects like, e.g., a possible expansion of the system before splitting. However, we want to estimate the mean velocity of the fragments for two opposite scenarios, under some specific –but rather realistic– assumptions. The two scenarios are: the binary splitting of a deformed compound nucleus, investigated at the scission configuration; the binary splitting of an undeformed compound nucleus into two touching spheres.

For the first scenarios we make use of the assumptions introduced in the statistical model of Wilkins, Steinberg and Chasman [81], where the total kinetic energy was considered determined basically by the coulomb repulsion of the two fragments at the scission point, whereas other terms are negligible, like for instance the energy from saddle to scission, which is mostly lost in dissipative phenomena. The mean velocity of the fission fragments is estimated by the following empirical liquid-drop description of the total kinetic energy:

$$TKE = \frac{Z_1 Z_2 e^2}{D} \quad \text{with} \quad D = r_0 A_1^{1/3}\left(1 + \frac{2\beta_1}{3}\right) + r_0 A_2^{1/3}\left(1 + \frac{2\beta_2}{3}\right) + d \quad (7)$$

where $A_1$, $A_2$, $Z_1$, $Z_2$ denote the mass and atomic numbers of a pair of fission fragments prior to neutron evaporation. $D$ represents the distance between the two charges and is given by the fragment radii ($r_0 A^{1/3}$), corrected for the deformation ($\beta$), plus the neck ($d$). The parameters ($r_0$=1.16 fm, $d$=2.0 fm, $\beta_1=\beta_2$=0.625) were deduced from experimental data in ref. [80] and are consistent with values previously found in the analysis of ref. [81]. The formula (7) is valid for sufficiently excited nuclei, where shell effects are negligible. When the momentum conservation is imposed to the reaction, the velocities of the two fission fragments are determined. We have estimated the mean velocities of the fission fragments for two compound nuclei: $^{238}_{92}U$ and $^{185}_{79}Au$. They are compared with our data in Fig. 13. While for the heavier fragments, the experimental data fall in between these two estimates, for fragments below $Z = 25$, the mean velocity tends to be higher than the estimation for the $^{238}U$ compound nucleus. This indicates that the experimental parameters of equation 7 that were obtained in symmetric fission are not applicable to very asymmetric mass splits. In very asymmetric fission, both the neck (parameter $d$) and the deformation (parameter $\beta$) seem to be smaller, with a consequent increase of the kinetic energy.

The opposite extreme of the situation described by equation 7 is the scenario of asymmetric binary decay from undeformed nuclei. In this context, we assumed that the binary decay can be described as the inverse process of fusion. The shape of the potential is given in terms of the nuclear, Coulomb and centrifugal contributions:

$$V_{eff} = -V_N + \frac{Z_1 \cdot Z_2 \cdot e^2}{r} + \frac{l(l+1) \cdot \hbar^2}{2\mu r^2} \quad (8)$$

where $Z_i$ are the charges, $r$ is the distance between the centers of the nuclei, $\mu$ is the reduced mass, and $l$ is the quantum number for the angular momentum. In our calculations, the empirical nuclear potential of R.Bass [82, 83] is used. The total kinetic energy of the two nuclei is assumed to be equal to the height of the fusion potential barrier. Imposing momentum conservation, the velocity of the two fragments was determined. The result of this calculation for the compound nuclei $^{238}_{92}U$ and $^{185}_{79}Au$ is represented in Fig. 13 by the dashed lines.

The comparison of the experimental data with the two set of calculations seems to indicate a tendency of going from a split into highly deformed nuclei to a split into undeformed nuclei as the charge of the fragments decreases. This result gives an indication that the lightest fragments are produced in configurations which are more compact than predicted by the systematics of equation 7 that is based on more symmetric fission.

The failure of the descriptions for the kinetic energies of very light fission fragments, which were deduced from symmetric fission of heavy systems, e.g. ref. [84], was already noticed and lead to some modified empirical formulations [85 ,20].

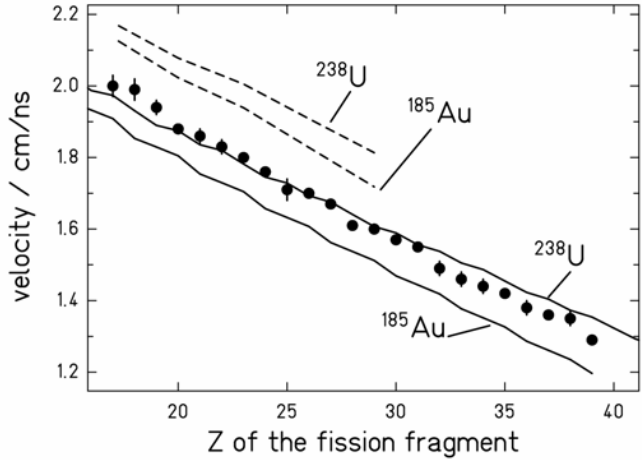

Fig. 13: Measured mean values of the velocities of fission fragments in the frame of the fissioning nucleus (●). The lines represent the expected values of the velocity of fragments originating from the compound nuclei $^{238}$U and $^{185}$Au. The solid lines represent the expected velocities for the scission-point model (deformed nuclei) and the dashed lines the values obtained by the nucleus-nucleus fusion approach (undeformed nuclei).

*6.2.3 Comparison with the ABRABLA code*

*Charge and mass distributions*

In the charge and mass distributions of Fig. 11, the data show a clear deviation from a Gaussian shape at about $Z \approx 18$, $A \approx 40$.

In the following, we want to show that the observed change of slope of the distributions can be explained by means of the statistical model, by the binary decay of a fully equilibrated compound nucleus. To this purpose, we inserted the BURST model [86] in the statistical abrasion-ablation code ABRABLA [87, 88, 89], so that also spallation-ablation reactions can be treated now. Both codes were developed at GSI.

In the analytical code BURST, the prefragments arising from high-energy nucleon-nucleus collisions are calculated. The code is based on the parameterisation of the output results of the intranuclear-cascade stage predicted by INCL3 [90]. It gives a consistent description of the numbers and the kinetic energies of protons and neutrons removed from the target and of the excitation energy and angular momentum acquired in the cascade of individual high-energy nucleon-nucleon collisions.

In the de-excitation stage of ABRABLA, named ABLA, the compound nucleus at every step of its evolution has two possible decay channels: evaporation and fission. Evaporation is treated as described in ref. [88], the determination of the fission yields as described in ref. [89] and the dynamical evolution of fission as described in ref. [91]. In the statistical model of fission for a given excitation energy the yield of a certain fission fragment is determined by the statistical weight of the transition states above the potential barrier, i.e. at the saddle point. This weight is in turn correlated to the density of nuclear levels. In ABLA, the latter are calculated using the thermodynamic Fermi-gas picture, i.e. assuming the nucleus as a system of non-interacting fermionic particles. The potential energy at the saddle point depends on mass-asymmetric deformations, which lead to the formation of two fragments of different sizes. In the fission model of ABLA, the barrier as a function of mass asymmetry is defined by three components. The first is the symmetric component defined by the liquid-drop potential by means of a parabolic function with a curvature obtained from experimental data [92]. This parabola is modulated by two neutron shells, represented by Gaussian functions. Shells are supposed to wash out with excitation energy [93]. The heights and the widths of the Gaussians representing the shell effects and additional fluctuations in mass asymmetry acquired from saddle to scission are derived from experimental data [89]. The above representation of the barrier as a function of mass asymmetry is valid only for the main fission region (from $Z \approx 20$ to $Z \approx 65$), while for very asymmetric mass splits the potential energy is expected to inverse the slope and start to decrease, as discussed in section 6.2.1. Up to now, this approximation was considered sufficient since fission was expected to die out rapidly below $Z \sim 28$.

In order to correctly describe the binary decay of the compound nucleus also for very asymmetric mass-splits one should properly model the M-shaped potential energy as a function of the mass asymmetry (Fig. 12-left). This is for instance the treatment applied in GEMINI [94]. We used another approach. In the previous section, we concluded that the lightest fragments are produced in a rather compact configuration. We take this evidence as an indication that there is gradual transition from the standard fission process towards evaporation. From the physical point of view an extremely asymmetric binary split into two compact nuclei corresponds to an evaporation of a light nucleus from a heavy compound nucleus. Up to now the evaporation part of ABLA considered only the emission of light particles, specifically: neutrons, protons, tritons, deuterons, $^3$He and alphas. In the code, we extended the evaporation to intermediate-mass fragments (IMF), i.e. to the emission of light nuclei with $Z>2$. The statistical weight for the emission of these fragments is calculated on the basis of the detailed-balance principle. The decay width ($\Gamma$) as a function of the excitation energy ($E$) depends on the inverse cross section ($\sigma_{inv}$), on the level densities of the two daughter nuclei ($\rho_{imf}$ and $\rho_{partner}$) and on the level density of the mother nucleus above the ground state ($\rho_C$):

$$\Gamma \approx \int_0^{E_{imf}^{max}} \int_0^{E_{partner}^{max}} \sigma_{inv} \frac{\rho_{imf}(E_{imf}) \cdot \rho_{partner}(E_{partner})}{\rho_C(E)} dE_{imf} \, dE_{partner} \qquad (9)$$

with the following relation that guaranties the energy conservation:

$$E = E_{imf} + E_{partner} + Q + \varepsilon - B \qquad (10)$$

Here $E$, $E_{imf}$ and $E_{partner}$ represent the initial excitation energy of the mother nucleus, and the excitation energies of the two daughter nuclei, respectively. Q is the Q-value, $\varepsilon$ is the total kinetic energy in the centre of mass of the system, and $B$ is the barrier of the potential. The barrier ($B$) was calculated using the fusion nuclear potential of Bass [82, 83]. The inverse cross section ($\sigma_{inv}$) was calculated using the ingoing-wave boundary condition model [95], where only a real potential is used to describe the transmission probability of particles. An analytical approximation to equation 9 was used in order to avoid the numerical calculation of the two integrals, which is rather time-consuming. This technical procedure will be described elsewhere.

One may object that, although the mean velocities of the fragments presented in Fig. 13 indicate that the two nuclei are formed in a rather compact configuration, they are not completely underformed. The deformation energy should be included to have a consistent description. On the other hand, other effects, like the thermal expansion of the excited nucleus, the surface effects on the level densities, the pre-formation probabilities, can affect the decay width. They influence the result in opposite ways, the ones increasing, the others decreasing the decay width. Considering the good agreement of our calculation with the experimental results (see later Fig. 14, Fig. 15 and Fig. 16), the global influence of all these contributions seems to be small.

In Fig. 14, the result of the code is presented for the entire production range both on the chart of the nuclides (up) and as a charge distribution (down). The latter is compared with the experimental data. The figure includes the heavier fragments obtained in the parallel analysis [11, 12, 13]. We recall that our measurement was technically limited to $Z \geq 7$, but the production of light nuclides would extend even farther down to $Z=1$. The full line is obtained by the sum of the three components: the evaporated IMF, the fission fragments and the heavy evaporation residues (the evaporated light-charged particles ($Z \leq 2$) are also evaluated by the code, but not included in the figure here). In Fig. 15, eight isotopic distributions are compared with the calculation. In all the comparisons, the agreement is very good, proving that the binary decay of a fully equilibrated compound nucleus contributes in a dominant way to the production of light fragments.

We would like to point out that subdividing the description of the binary decay in two parts (IMF emission and standard fission) has the advantage of bypassing the description of the M-shaped conditional saddle, which is not an easy task. On the contrary, the semiempirical approach used in ABLA has proven to be versatile and to have a very good predictive power, especially for the description of low-energy fission, where the modelling of the fission channels play a decisive role [89].

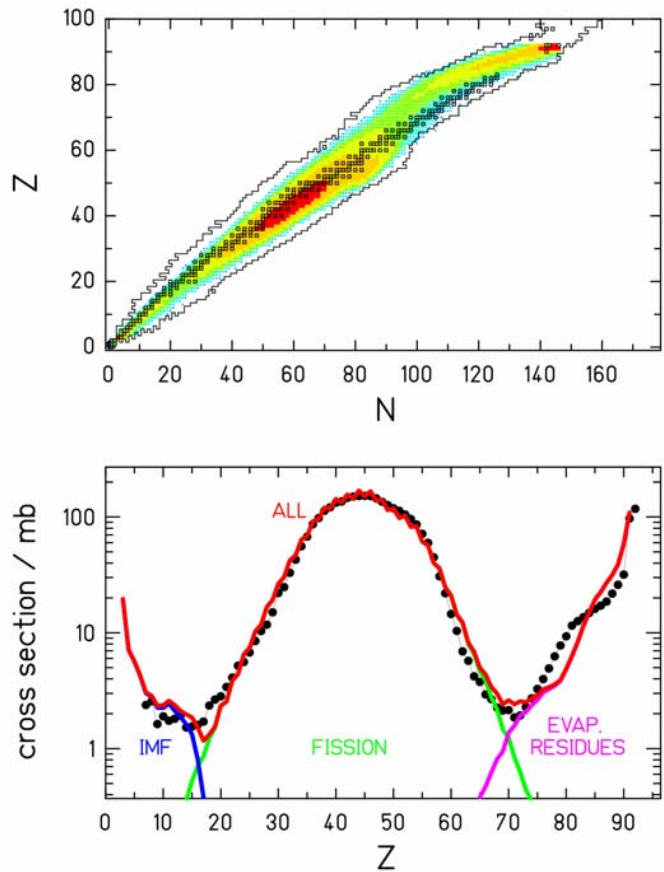

Fig. 14: (Colour on-line) Cross sections for the nuclei produced in $^{238}$U (1 $A$ GeV) + p. Up: Prediction of ABRABLA presented on the chart of the nuclides. Down: Experimental data (full dots) [this work, 11, 12, 13] are compared with the results of ABRABLA (solid line). The solid line is obtained by the sum of the three components: the evaporated IMF, the fission fragments and the heavy evaporation residues (dashed lines).

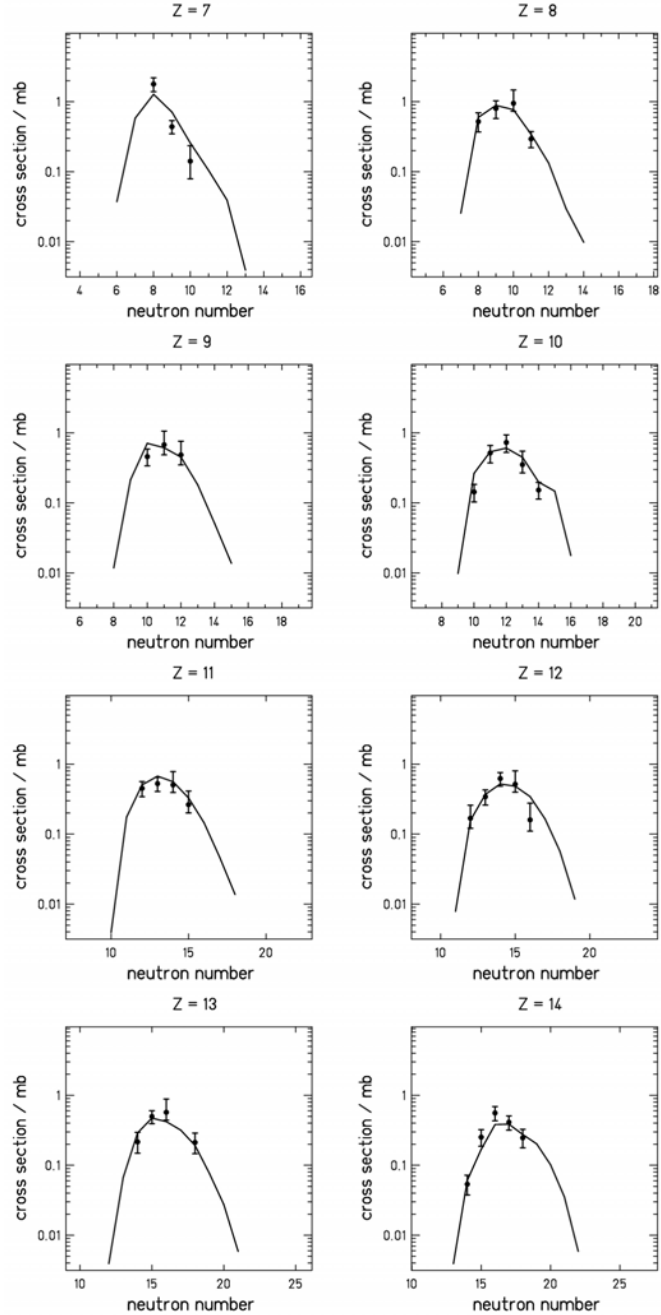

Fig. 15: Cross sections for the isotopes of the eight lightest elements measured in the reaction $^{238}$U (1 *A* GeV) + p. The dots represent the experimental data, measured in this work, and the solid lines the prediction of ABRABLA, which includes the evaporation of IMFs.

*Mean value and width of the isotopic distributions*

In Fig. 16, the mean *N/Z*-ratio and the standard deviations of the isotopic distributions are shown as a function of the atomic number for the entire production, which includes also the data of ref. [11, 12, 13]. The dashed line represents the stable isotopes, the solid line the result of the ABRABLA prediction.

In the fission model inside the ABLA code, the population of the fission channels is assumed to be basically determined by the statistical weight of transition states above the potential-energy landscape at the fission barrier, as described previously. Several properties, however, are finally determined at scission, among them the mean value and the fluctuations in the neutron-to-proton ratio, which are responsible for the so-called "charge polarisation" [96]. The fluctuations in the neutron-to-proton ratio are considered by describing the potential in this degree of freedom by a parabolic function. Assuming that the equilibration in this variable is fast compared to the saddle-to-scission time, the curvature of this potential was calculated in a touching-sphere configuration. From the knowledge of both the excitation energy and deformation energy of the system at the scission point, the excitation energies of the two fission pre-fragments can be sampled. The final fission fragments are then obtained at the end of the respective evaporation cascades. A full description of the model is given in ref. [89].

It can be noticed that the ABRABLA calculation reproduces correctly the mean values (the $<N>/Z$-ratio) of the isotopic distributions. Also the widths are well described, as can be noticed in Fig. 16. This is an indication that both the charge polarisation in the fission process and the competition with the evaporation of nucleons in the statistical model are rather well described in the code.

The light products are neutron rich, as expected to be in fission. Compared to electromagnetic-induced fission (see for instance ref. [37, 39, 40]), where the mean $N/Z$ is closer to that one of $^{238}$U, here the neutron excess is lower, demonstrating that the process occurred at higher excitation energies. The neutron enrichment decreases slightly with the decreasing mass, as well as the width of the distribution. The latter effect is more evident. The reason for these tendencies is connected to the fact that the valley of stability becomes quickly narrow and steep. Large fluctuations in $N/Z$ become more and more unlikely.

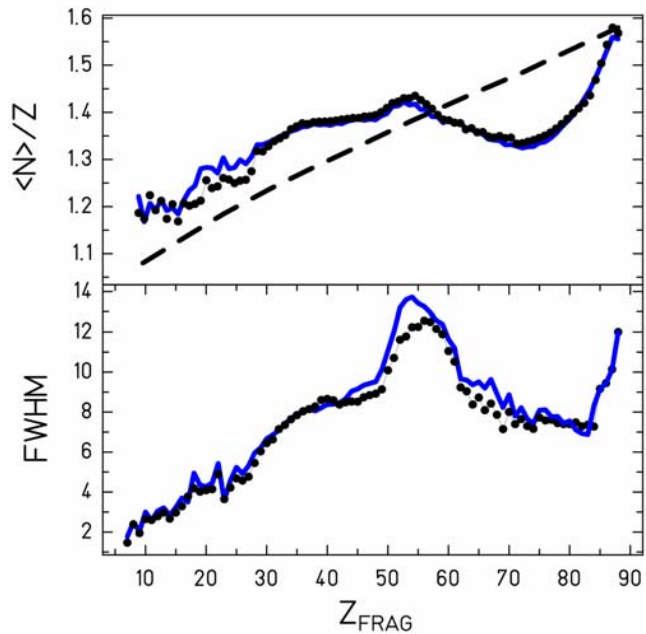

Fig. 16: Left: Mean neutron-to-proton ratio of isotopic distributions as a function of the atomic number, compared with the stability line (dashed line) and to the ABRABLA prediction (solid line). Right: FWHM of the isotopic distributions compared to the prediction of the ABRABLA code (solid line).

To conclude this section, we like to point out that the result of the ABRABLA code is remarkable, because the theoretical model behind it could never be compared with experimental results on fully identified nuclide distributions in the region of light fission fragments from proton-induced fission before. This is important, not only for the physics content, but also for the technical applications, where most of the available codes used to predict formation cross sections in fission reactions are based on empirical systematics (e.g. [97]), whose predictive power has proven to be rather poor [98].

*6.2.4 Ternary fission*

In the previous section we showed that all the experimental observables are consistent with the characteristics expected for fragments produced in binary fission. On the other hand, since only one particle was recorded in one event, one may question whether the observed light fragments could be produced in ternary fission [21]. In this context, it is useful to recall the results presented in ref. [99], where the kinetic-energy distributions of Be and C fragments observed in coincidence with fission fragments are presented. The light Be and C nuclei emitted at 50º with respect to the scission axis, present Gaussian-like distributions peaked at around 43 and 53 MeV, respectively. However, when the light Be and C nuclei are observed at 90º, two additional Gaussian-like humps appear at 22 MeV and 26 MeV for Be and C, respectively. The authors interpret these data assuming that the component at higher energy is associated with pre-scission emission events, which are essentially isotropic. These nuclei are presumably emitted in an earlier step of the deexcitation cascade preceding the fission process. The component at lower energy is due to ternary fission, i.e. to light nuclei emerging from the neck of the system, travelling at about 90º with respect to the scission axis.

In the present experiment, the existence of a lower-energy component would result in an additional concentric shell with smaller radius in the kinematical pattern schematically represented in Fig. 4, and consequently in five-fold humped velocity spectra. Such an additional lower-energy hump was never observed in the velocity spectra of our data (see Fig. 5). Therefore, we conclude that the light nuclei presented here were not produced in ternary fission.

*6.3 Fast break-up process*

In the preceding sections we have investigated the contribution of fission. We have concluded that fission plays a major role in the production of light fragments. However, it could be not the only process responsible for the formation of light products. Here we want to discuss the possible contribution of a fast break-up process. With fast break-up we mean a multifragmentation-like reaction mechanism. The dynamical picture, thought to be behind it, is that of a *fast* thermal expansion right after the intra-nuclear cascade phase, with the formation of two clusters, successively driven apart by the Coulomb repulsion. In contrast to this fast process, the fission-evaporation picture assumes the *slow* decay of a fully equilibrated compound nucleus. Observables, which could hint at one or the other process, are the time scale of the process, the multiplicity distribution of the products, the excitation energy of the decaying system, its momentum transfer, and the mean velocity of the fragments. In the following, we will analyse all these signatures. We will also critically investigate the justification of previously drawn conclusions in some other works that found indications for fast break-up processes in similar systems.

*Multiplicity*

In our experiment, the measurement of multiplicity was not possible. Very light fragments from lithium to argon were also investigated in 1 GeV proton-induced spallation of gold and some lighter nuclei [30]. It was observed that the probability for multiple IMF production ($Z \geq 3$) with a multiplicity $\geq 3$ in the reaction Au + $^1$H at 1 GeV is only 0.4 % (i.e. almost all decays are binary). One may expect that in the system $^{238}$U + $^1$H at 1 $A$ GeV the percentage will be comparable, since the energies introduced and the sizes of the two systems are rather similar.

*Time scale*

A recent theoretical work [100] investigates the compatibility of the measured properties of light fragmentation products with a binary sequential decay model. They find that the experimental charge and energy distributions of the fragments produced in the spallation of gold by 8.1 GeV protons are well reproduced. Only the time scale, deduced from angular correlations of IMFs, is off by about a factor of three. Unfortunately, the method of angular correlations of light fragments is not applicable for the appreciably lower projectile energy of 1 GeV used in the present work, since the probability for the emission of more than one light fragment is very low [101]. Eventual evidence on the compatibility with a fission process in the generalized sense might only be drawn from the other observables.

*Excitation energy*

In the reaction Au + $^1$H at 1 GeV investigated in ref. [30], along with the multiplicity equal to 2, the velocity spectra of the emitted light fragments indicate that they are produced by the binary decay of a heavy nucleus. Thus, the production of light fragments in this reaction is predominantly a binary process, forming one heavy and one light fragment. In ref. [30], in order to deduce the excitation energy of the decaying system, the energy spectra were fitted with a Maxwellian distribution, assuming isotropic emission. The deduced slope parameter gave an apparent temperature of about 8.4 MeV for all light fragments between $Z = 7$ and $Z = 18$. If interpreted as a temperature value, this would correspond to an excitation energy of about 1400 MeV in a fully thermalised system, which is even higher than the centre-of-mass energy, available in the reaction. However, in our opinion, the slope parameter cannot be interpreted as the temperature of the emitting source, because it is strongly influenced by several additional effects. One is the Fermi motion of the nucleons in a nucleus which is breaking up. This effect has been described by Goldhaber in [43]. Its relevance for the interpretation of the kinetic properties of nuclear decay products has been underlined by Westfall et al. [70] and recently discussed in [102]. That means that the slope parameter of the energy spectra of the light fragments observed by [30] mostly reflects the velocity distribution of the nucleons in the decaying system and thus cannot be attributed to the characteristics of the decay process. There is also another effect, which has an important influence on the interpretation of the energy spectra. It relates to the fact that the light fragments may be produced by the decay of a variety of mother nuclei with different mass and atomic number. Also this fact was not considered in [30]. This effect alone causes an important fluctuation on the kinetic energy of the emitted fragments. The two effects discussed, the Fermi motion and the variety of emitting sources, make it rather difficult to find a straight-forward quantitative interpretation of the slope parameter in term of "temperature parameter", in the energy spectra of the IMFs produced in a 1 GeV proton-induced spallation reaction. We conclude that the large value of the slope parameter cannot be taken as a proof for a fast binary decay, occurring before the formation of a thermalised compound nucleus as done in ref. [30].

*Momentum transfer*

Barz et al. [31] reported folding-angle distributions of binary-decay products from the spallation of uranium, samarium and silver by 1 GeV protons. While for the binary-decay products of uranium the momentum transfer and its fluctuation are small, both quantities increase when going to samarium products. Fragments produced in the spallation of silver reveal a very large spread in momentum transfer, but no further increase of the momentum transfer. These findings were interpreted as an indication for the onset of multifragmentation in the lighter systems. As for the excitation energy, also the very large spread in momentum transfer in the spallation of silver [31] can at least partly be related to the fact that the light fragments may be produced by the fission of a variety of mother nuclei with different mass and atomic number, without the need of introducing a multifragmentation process. Also the Fermi motion of the abraded nucleons produces a similar effect. As discussed in ref. [12], the same argument might also explain great part of the rather broad relative kinetic energy width found in the present work (around 18 % standard deviation, see Tab. 1) if compared to the energy width known from low-energy fission of uranium isotopes, which amounts to about 5% only [103].

*Mean velocities of the fragments*

The mean velocity of the fragments with respect to the emitting source was the key information from which the binary nature of the decay was deduced. In Fig. 13 the comparison of the data with the results of calculations performed assuming a fission-evaporation scenario seems to reproduce satisfactorily the data. Under the hypothesis of a fast binary break-up, the expansion stage would result in a larger distance between the two clusters, with the consequence of a reduction of the Coulomb repulsion and eventually of the mean velocity. The possible presence of a third small cluster would sort out the same effect. In the end, we conclude that the mean velocities represent a rather strong evidence that the reaction mechanism is a generalised fission process.

A similar investigation was performed already in 1987 by Andronenko [65]. He analysed all the signatures (among which angular correlations, mass and energy distributions) of the binary products from several proton-induced reactions at 1 GeV available at that time. The interaction of a proton with nuclei followed by fission, described applying a cascade-evaporation model, could reproduce all the observed signatures, and he excluded the contribution of other decay modes. Similar conclusion were also drawn by Jahnke et al. [104], studying the binary decay of uranium from the antiproton-induced reactions at 1 GeV. Also Lott et al. [105] recently confirmed that binary products of uranium from the antiproton-induced reactions at 1.22 GeV do not show any signature of multifragmentation.

We conclude that the results from several experiments, including the present work, give unanimous indications that light fragments in the reaction $^{238}$U + $^{1}$H at 1 $A$ GeV are produced in a binary decay. Although the nature of this decay could not be identified without doubt, clear indications for a fast break-up process in this reaction seem to be absent. On the contrary, it may be concluded that, at the current stage of knowledge, the experimental signatures in the reaction $^{238}$U + $^{1}$H at 1 $A$ GeV are consistent with the binary decay of a fully equilibrated compound nucleus.

## 7. Conclusions

Despite the long study of fission, there is still very little experimental information available on the light residues produced in the fission of actinides. Here, 254 light residues in the element range 7≤Z≤37 formed in the proton-induced reactions of $^{238}$U at 1 $A$ GeV were presented. This experimental work belongs to a systematic study of the reaction $^{238}$U + $^{1}$H at 1 $A$ GeV, where the production of nuclides with 7≤Z≤92 was measured and analysed. The other experimental data, complementing those presented here, can be found in refs. [11, 12, 13].

The light fragments presented here, which populate the chart of the nuclides far down, could be qualified as binary-decay products thanks to the available kinematic information. A detailed study of all the experimental observables – the mass and charge distributions, the isotopic distributions, the mean velocities, the width of the velocity distributions, the mean recoil velocities of the mean mother nuclei – showed that all the above-quoted signatures are consistent with the binary decay of a fully equilibrated compound nucleus, while clear indications for fast break-up processes seem to be absent. As discussed in [78], the binary decay of a compound nucleus includes fission and evaporation with a natural transition in-between, and it might be called fission in a generalized sense [73]. Thus, very asymmetric fission of the system $^{238}$U + $^{1}$H at 1 $A$ GeV seems to reach down to rather light nuclei, extending below $Z = 7$ and merge with evaporation. In the spallation-fission reaction of $^{238}$U this feature is unambiguously identified for the first time.

**Acknowledgments**

We wish to thank F. Ameil, K. Günzert and M. Pravikoff for their participation to the data taking, K. H. Behr, A. Brünle and K. Burkard for their technical support during the experiment and the group of P. Chesny for building and operating the liquid-hydrogen target. The work has been supported by the European Community with the programmes "Human Capital and Mobility" under the contract ERBCHB CT 94 0717, "Access to Research Infrastructure Action of the Improving Human Potential" under the contract HPRI 1999 CT 00001, "HINDAS" under the contract FIKW CT 2000 00031, and "EURISOL" under the contract HPRI-1999-CT-50001.

# Appendix

Table A.1: Measured fission cross sections for the spallation of 1 $A$ GeV $^{238}$U on hydrogen. The last two columns represent the upper and lower relative uncertainties (expressed in percentage). Both statistical and systematic errors are considered.

A complete set of data for the reaction 1 $A$ GeV $^{238}$U on protons [this work, 11, 12, 13], are collected in [49]. There, the values for $Z<28$ are taken from the present work, the values for the isotopes of the elements $Z=28$ and $Z=29$ were obtained from the combination of the two experimental results obtained in this work and in ref. [12], while for the nuclei above $Z=30$ and $Z=37$ the data of ref. [12] are presented.

| $Z$ | $N$ | $\sigma$/mb | $\varepsilon_{rel}^{up}$ | $\varepsilon_{rel}^{dw}$ |
|---|---|---|---|---|
| 7 | 8 | 1.8 | 23 | 22 |
| 7 | 9 | 0.44 | 22 | 21 |
| 7 | 10 | 0.14 | 67 | 44 |
| 8 | 8 | 0.52 | 33 | 29 |
| 8 | 9 | 0.8 | 28 | 28 |
| 8 | 10 | 0.95 | 55 | 23 |
| 8 | 11 | 0.29 | 27 | 25 |
| 9 | 10 | 0.46 | 28 | 26 |
| 9 | 11 | 0.68 | 57 | 28 |
| 9 | 12 | 0.49 | 57 | 28 |
| 10 | 10 | 0.14 | 28 | 28 |
| 10 | 11 | 0.52 | 28 | 28 |
| 10 | 12 | 0.73 | 28 | 28 |
| 10 | 13 | 0.36 | 56 | 24 |
| 10 | 14 | 0.15 | 29 | 26 |
| 11 | 12 | 0.45 | 25 | 24 |
| 11 | 13 | 0.53 | 23 | 23 |
| 11 | 14 | 0.5 | 55 | 22 |
| 11 | 15 | 0.26 | 56 | 24 |
| 12 | 12 | 0.17 | 54 | 28 |
| 12 | 13 | 0.34 | 25 | 24 |
| 12 | 14 | 0.62 | 22 | 22 |
| 12 | 15 | 0.52 | 55 | 23 |
| 12 | 16 | 0.16 | 73 | 31 |
| 13 | 14 | 0.22 | 37 | 31 |
| 13 | 15 | 0.5 | 21 | 21 |
| 13 | 16 | 0.57 | 55 | 23 |
| 13 | 17 | == | | |
| 13 | 18 | 0.21 | 36 | 31 |
| 14 | 14 | 0.054 | 34 | 30 |
| 14 | 15 | 0.25 | 28 | 26 |
| 14 | 16 | 0.56 | 23 | 23 |
| 14 | 17 | 0.41 | 23 | 23 |
| 14 | 18 | 0.25 | 32 | 28 |
| 15 | 16 | 0.12 | 29 | 27 |
| 15 | 17 | 0.39 | 21 | 21 |
| 15 | 18 | 0.47 | 28 | 28 |
| 15 | 19 | 0.34 | 57 | 28 |
| 15 | 20 | 0.23 | 56 | 24 |
| 16 | 17 | 0.11 | 29 | 27 |
| 16 | 18 | 0.4 | 22 | 22 |
| 16 | 19 | 0.46 | 28 | 28 |
| 16 | 20 | 0.35 | 54 | 21 |
| 16 | 21 | 0.2 | 57 | 26 |
| 16 | 22 | 0.084 | 33 | 29 |
| 17 | 18 | 0.18 | 32 | 28 |
| 17 | 19 | 0.29 | 24 | 23 |
| 17 | 20 | 0.4 | 28 | 26 |
| 17 | 21 | 0.37 | 55 | 23 |
| 17 | 22 | 0.29 | 28 | 28 |
| 17 | 23 | 0.18 | 55 | 23 |
| 18 | 19 | 0.11 | 35 | 30 |
| 18 | 20 | 0.24 | 23 | 22 |
| 18 | 21 | 0.55 | 25 | 24 |
| 18 | 22 | 0.59 | 22 | 22 |
| 18 | 23 | == | | |
| 18 | 24 | 0.27 | 55 | 22 |
| 18 | 25 | 0.096 | 61 | 58 |
| 18 | 26 | 0.047 | 44 | 35 |
| 19 | 20 | 0.041 | 38 | 32 |
| 19 | 21 | 0.15 | 24 | 23 |
| 19 | 22 | 0.33 | 27 | 25 |
| 19 | 23 | 0.57 | 22 | 22 |
| 19 | 24 | 0.58 | 56 | 23 |
| 19 | 25 | 0.62 | 57 | 26 |
| 19 | 26 | 0.23 | 57 | 25 |
| 19 | 27 | 0.13 | 43 | 34 |
| 20 | 21 | 0.057 | 29 | 27 |
| 20 | 22 | 0.18 | 23 | 23 |
| 20 | 23 | 0.39 | 25 | 24 |
| 20 | 24 | 0.61 | 21 | 21 |
| 20 | 25 | 0.54 | 21 | 21 |
| 20 | 26 | 0.64 | 24 | 23 |
| 20 | 27 | 0.32 | 23 | 23 |
| 20 | 28 | 0.091 | 66 | 34 |
| 21 | 25 | 0.5 | 22 | 22 |
| 21 | 26 | 0.76 | 21 | 21 |
| 21 | 27 | 0.59 | 28 | 28 |
| 21 | 28 | 0.65 | 21 | 21 |
| 21 | 29 | == | | |
| 21 | 30 | 0.07 | 59 | 28 |
| 21 | 22 | 0.04 | 36 | 30 |
| 21 | 23 | 0.15 | 26 | 25 |
| 21 | 24 | 0.38 | 23 | 23 |
| 22 | 24 | 0.11 | 23 | 22 |
| 22 | 25 | 0.39 | 24 | 23 |
| 22 | 26 | == | | |
| 22 | 27 | 0.9 | 20 | 20 |
| 22 | 28 | == | | |
| 22 | 29 | 0.71 | 21 | 21 |
| 22 | 30 | 0.29 | 57 | 26 |
| 22 | 31 | 0.13 | 58 | 27 |
| 22 | 32 | 0.03 | 60 | 52 |
| 23 | 25 | 0.086 | 28 | 28 |
| 23 | 26 | 0.36 | 22 | 22 |
| 23 | 27 | == | | |
| 23 | 28 | 1.1 | 20 | 20 |
| 23 | 29 | 1.3 | 57 | 28 |
| 23 | 30 | 1.1 | 28 | 28 |
| 23 | 31 | 0.55 | 23 | 22 |
| 24 | 26 | 0.062 | 23 | 22 |
| 24 | 27 | 0.34 | 22 | 21 |
| 24 | 28 | 0.59 | 24 | 23 |
| 24 | 29 | 1 | 20 | 20 |
| 24 | 30 | 1.1 | 57 | 28 |
| 24 | 31 | 1.3 | 22 | 22 |
| 24 | 32 | 0.73 | 21 | 20 |
| 24 | 33 | == | | |
| 24 | 34 | 0.18 | 24 | 23 |
| 25 | 27 | 0.047 | 25 | 24 |
| 25 | 28 | 0.28 | 22 | 22 |
| 25 | 29 | == | | |
| 25 | 30 | 1.3 | 21 | 21 |
| 25 | 31 | 1.3 | 21 | 21 |
| 25 | 32 | == | | |
| 25 | 33 | 0.95 | 20 | 20 |
| 25 | 34 | 0.53 | 24 | 23 |
| 25 | 35 | == | | |
| 25 | 36 | 0.11 | 58 | 26 |
| 26 | 27 | 0.0082 | 28 | 28 |
| 26 | 28 | 0.045 | 24 | 23 |
| 26 | 29 | 0.26 | 22 | 22 |
| 26 | 30 | 0.59 | 24 | 23 |
| 26 | 31 | 1.3 | 21 | 21 |
| 26 | 32 | 1.6 | 57 | 28 |
| 26 | 33 | 1.4 | 28 | 28 |
| 26 | 34 | 1.6 | 20 | 20 |
| 26 | 35 | 1 | 28 | 28 |
| 26 | 36 | == | | |
| 26 | 37 | 0.23 | 23 | 22 |
| 27 | 29 | 0.039 | 24 | 23 |
| 27 | 30 | 0.23 | 22 | 22 |
| 27 | 31 | == | | |
| 27 | 32 | 1.2 | 21 | 21 |
| 27 | 33 | 1.6 | 20 | 20 |
| 27 | 34 | 1.7 | 28 | 28 |
| 27 | 35 | 2.2 | 20 | 20 |
| 27 | 36 | 1.8 | 28 | 28 |
| 27 | 37 | == | | |
| 27 | 38 | 0.55 | 54 | 21 |
| 27 | 39 | 0.24 | 58 | 27 |
| 28 | 30 | 0.028 | 26 | 24 |
| 28 | 31 | 0.11 | 22 | 22 |
| 28 | 32 | 0.31 | 57 | 25 |
| 28 | 33 | 0.87 | 22 | 21 |
| 28 | 34 | 1.6 | 20 | 20 |
| 28 | 35 | 2.1 | 28 | 28 |
| 28 | 36 | 2.3 | 21 | 21 |
| 28 | 37 | 2.3 | 28 | 28 |
| 28 | 38 | 1.8 | 28 | 28 |
| 28 | 39 | == | | |
| 28 | 40 | == | | |
| 28 | 41 | == | | |
| 28 | 42 | == | | |
| 28 | 43 | 0.01 | 54 | 39 |
| 28 | 44 | 0.035 | 24 | 23 |
| 28 | 45 | 0.0058 | 50 | 37 |
| 29 | 31 | 0.011 | 30 | 27 |
| 29 | 32 | 0.075 | 24 | 23 |
| 29 | 33 | 0.3 | 28 | 28 |
| 29 | 34 | == | | |
| 29 | 35 | 1.5 | 20 | 20 |
| 29 | 36 | 2.2 | 28 | 28 |
| 29 | 37 | 2.7 | 21 | 21 |
| 29 | 38 | 2.8 | 20 | 20 |
| 29 | 39 | 2.3 | 28 | 28 |
| 29 | 40 | 1.4 | 24 | 23 |
| 29 | 41 | == | | |
| 29 | 42 | == | | |
| 29 | 43 | == | | |
| 29 | 44 | == | | |
| 29 | 45 | 0.084 | 24 | 23 |
| 29 | 46 | 0.03 | 26 | 25 |
| 29 | 47 | 0.0041 | 36 | 30 |
| 30 | 32 | 0.014 | 32 | 28 |
| 30 | 33 | 0.059 | 22 | 22 |
| 30 | 34 | == | | |
| 30 | 35 | 0.61 | 28 | 28 |
| 30 | 36 | 1.1 | 21 | 21 |
| 30 | 37 | 2.4 | 57 | 28 |
| 30 | 38 | 2.9 | 28 | 28 |
| 30 | 39 | 3.6 | 20 | 20 |
| 30 | 40 | 3.7 | 28 | 28 |
| 30 | 41 | 2.8 | 28 | 28 |
| 30 | 42 | == | | |
| 30 | 43 | == | | |
| 30 | 44 | == | | |
| 30 | 45 | == | | |
| 30 | 46 | == | | |
| 30 | 47 | 0.12 | 22 | 22 |
| 31 | 33 | 0.0039 | 38 | 31 |
| 31 | 34 | 0.047 | 24 | 23 |
| 31 | 35 | == | | |
| 31 | 36 | 0.48 | 21 | 21 |
| 31 | 37 | 1.1 | 54 | 21 |
| 31 | 38 | 2.3 | 57 | 28 |
| 31 | 39 | 3 | 28 | 28 |
| 31 | 40 | 4.4 | 20 | 20 |
| 31 | 41 | 3.6 | 21 | 20 |
| 31 | 42 | 3.8 | 28 | 28 |
| 31 | 43 | == | | |
| 31 | 44 | == | | |
| 31 | 45 | == | | |
| 31 | 46 | == | | |
| 31 | 47 | == | | |
| 31 | 48 | 0.19 | 21 | 21 |
| 31 | 49 | 0.16 | 22 | 22 |
| 31 | 50 | 0.025 | 31 | 23 |
| 32 | 35 | 0.034 | 23 | 23 |
| 32 | 36 | == | | |
| 32 | 37 | 0.41 | 28 | 28 |
| 32 | 38 | 1.1 | 21 | 21 |
| 32 | 39 | 1.7 | 22 | 22 |
| 32 | 40 | 3.4 | 28 | 28 |
| 32 | 41 | 4.7 | 21 | 21 |
| 32 | 42 | 4.5 | 20 | 20 |
| 32 | 43 | 5.1 | 28 | 28 |
| 32 | 44 | 3.4 | 21 | 21 |
| 32 | 45 | == | | |
| 32 | 46 | == | | |
| 32 | 47 | == | | |
| 32 | 48 | == | | |
| 32 | 49 | 0.73 | 59 | 41 |
| 32 | 50 | 0.39 | 28 | 28 |
| 32 | 51 | == | | |
| 32 | 52 | 0.06 | 28 | 28 |
| 33 | 36 | 0.013 | 26 | 25 |
| 33 | 37 | == | | |
| 33 | 38 | 0.35 | 22 | 22 |
| 33 | 39 | 0.89 | 21 | 21 |
| 33 | 40 | 1.8 | 21 | 21 |
| 33 | 41 | 2.9 | 57 | 28 |
| 33 | 42 | 5.1 | 21 | 21 |
| 33 | 43 | 4.6 | 29 | 26 |
| 33 | 44 | 6.5 | 28 | 28 |
| 33 | 45 | 5.6 | 28 | 28 |
| 33 | 46 | == | | |
| 33 | 47 | == | | |
| 33 | 48 | == | | |
| 33 | 49 | == | | |
| 33 | 50 | == | | |
| 33 | 51 | == | | |
| 33 | 52 | 0.62 | 21 | 21 |
| 34 | 37 | 0.017 | 28 | 28 |
| 34 | 38 | == | | |
| 34 | 39 | 0.26 | 22 | 22 |
| 34 | 40 | 0.82 | 22 | 21 |
| 34 | 41 | 1.2 | 21 | 21 |
| 34 | 42 | 3.1 | 28 | 28 |
| 34 | 43 | 4.7 | 28 | 28 |
| 34 | 44 | 5.7 | 20 | 20 |
| 34 | 45 | 8.6 | 28 | 28 |
| 34 | 46 | 7.2 | 28 | 28 |
| 34 | 47 | 5.4 | 21 | 21 |
| 34 | 48 | == | | |
| 34 | 49 | == | | |
| 34 | 50 | == | | |
| 34 | 51 | == | | |
| 34 | 52 | 0.89 | 65 | 43 |
| 34 | 53 | 1.2 | 21 | 21 |
| 34 | 54 | 0.77 | 24 | 23 |
| 35 | 38 | 0.0098 | 31 | 28 |
| 35 | 39 | 0.03 | 57 | 28 |
| 35 | 40 | 0.13 | 25 | 24 |
| 35 | 41 | 0.16 | 22 | 21 |
| 35 | 42 | 1.1 | 21 | 21 |
| 35 | 43 | 2.2 | 57 | 28 |
| 35 | 44 | 4.6 | 28 | 28 |
| 35 | 45 | 6.3 | 20 | 20 |
| 35 | 46 | 6.9 | 32 | 28 |
| 35 | 47 | 7.8 | 28 | 28 |
| 35 | 48 | 8.8 | 28 | 28 |
| 35 | 49 | == | | |
| 35 | 50 | == | | |
| 35 | 51 | == | | |
| 35 | 52 | == | | |
| 35 | 53 | == | | |
| 35 | 54 | 1.4 | 44 | 30 |
| 35 | 55 | 1.3 | 22 | 22 |
| 35 | 56 | 0.3 | 22 | 22 |
| 35 | 57 | 0.034 | 22 | 22 |
| 36 | 39 | 0.0065 | 62 | 42 |
| 36 | 40 | 0.015 | 28 | 28 |
| 36 | 41 | 0.058 | 33 | 29 |
| 36 | 42 | 0.49 | 21 | 21 |
| 36 | 43 | 0.79 | 21 | 21 |
| 36 | 44 | 2.2 | 26 | 25 |
| 36 | 45 | 4.1 | 57 | 28 |
| 36 | 46 | 6.7 | 21 | 21 |
| 36 | 47 | 8.2 | 55 | 56 |
| 36 | 48 | 11 | 28 | 28 |
| 36 | 49 | 10 | 28 | 28 |
| 36 | 50 | == | | |
| 36 | 51 | == | | |
| 36 | 52 | == | | |
| 36 | 53 | == | | |
| 36 | 54 | == | | |
| 36 | 55 | == | | |
| 36 | 56 | 2 | 21 | 21 |
| 36 | 57 | 1.5 | 21 | 21 |
| 36 | 58 | 0.25 | 28 | 28 |
| 36 | 59 | 0.083 | 23 | 22 |
| 37 | 41 | 0.017 | 28 | 28 |
| 37 | 42 | == | | |
| 37 | 43 | 0.27 | 22 | 21 |
| 37 | 44 | 0.58 | 21 | 21 |
| 37 | 45 | 1.5 | 25 | 24 |
| 37 | 46 | 3.7 | 57 | 28 |
| 37 | 47 | 7 | 28 | 28 |
| 37 | 48 | 7 | 20 | 20 |
| 37 | 49 | 12 | 28 | 28 |
| 37 | 50 | 11 | 28 | 28 |
| 37 | 51 | 11 | 21 | 21 |
| 37 | 52 | == | | |
| 37 | 53 | 5.8 | 36 | 28 |
| 37 | 54 | == | | |
| 37 | 55 | == | | |
| 37 | 56 | == | | |
| 37 | 57 | == | | |
| 37 | 58 | 1.8 | 22 | 21 |
| 37 | 59 | 1 | 22 | 22 |
| 37 | 60 | 0.16 | 23 | 22 |
| 37 | 61 | 0.016 | 23 | 23 |

Table A.2: Measured mean velocities for the fission fragments from the spallation of 1 $A$ GeV $^{238}$U on hydrogen.

| $Z$ of fission fragment | Mean recoil velocity of mother nuclei in the beam frame (cm/ns) | Mean velocity in mother-nucleus frame (cm/ns) | Mean kinetic energy in the mother-nucleus frame (MeV) |
|---|---|---|---|
| 17 | -0.13 ± 0.02 | 2.00 ± 0.03 | 77.9 ± 0.8 |
| 18 | -0.09 ± 0.02 | 1.99 ± 0.03 | 82.4 ± 1.2 |
| 19 | -0.13 ± 0.01 | 1.94 ± 0.02 | 83.4 ± 1.3 |
| 20 | -0.12 ± 0.01 | 1.88 ± 0.01 | 82.0 ± 1.4 |
| 21 | -0.12 ± 0.01 | 1.86 ± 0.02 | 84.8 ± 1.6 |
| 22 | -0.13 ± 0.01 | 1.83 ± 0.02 | 86.1 ± 0.6 |
| 23 | -0.13 ± 0.01 | 1.80 ± 0.01 | 87.2 ± 1.3 |
| 24 | -0.12 ± 0.01 | 1.76 ± 0.01 | 86.8 ± 1.2 |
| 25 | -0.10 ± 0.01 | 1.71 ± 0.03 | 85.9 ± 1.2 |
| 26 | -0.12 ± 0.01 | 1.70 ± 0.01 | 87.9 ± 1.0 |
| 27 | -0.10 ± 0.01 | 1.67 ± 0.01 | 88.8 ± 1.3 |
| 28 | -0.12 ± 0.01 | 1.61 ± 0.01 | 86.9 ± 0.9 |
| 29 | -0.11 ± 0.01 | 1.60 ± 0.01 | 89.3 ± 1.4 |
| 30 | -0.11 ± 0.01 | 1.57 ± 0.01 | 88.8 ± 0.5 |
| 31 | -0.10 ± 0.01 | 1.55 ± 0.01 | 90.6 ± 1.4 |
| 32 | -0.12 ± 0.01 | 1.49 ± 0.02 | 86.2 ± 1.2 |
| 33 | -0.12 ± 0.01 | 1.46 ± 0.02 | 86.3 ± 1.5 |
| 34 | -0.12 ± 0.01 | 1.44 ± 0.02 | 86.3 ± 1.4 |
| 35 | -0.11 ± 0.01 | 1.42 ± 0.01 | 86.9 ± 1.2 |
| 36 | -0.12 ± 0.01 | 1.38 ± 0.02 | 84.9 ± 1.2 |
| 37 | -0.12 ± 0.01 | 1.36 ± 0.01 | 84.4 ± 1.4 |
| 38 | -0.11 ± 0.01 | 1.35 ± 0.02 | 77.9 ± 0.8 |
| 39 | -0.12 ± 0.01 | 1.29 ± 0.01 | 82.4 ± 1.2 |